\documentclass[%
 reprint,
 amsmath,amssymb,
 aps,
]{revtex4-2}

\DeclareUnicodeCharacter{2008}{\,}

\usepackage{graphicx}
\usepackage{dcolumn}
\usepackage{bm}
\usepackage{hyperref}
\usepackage{color}
\usepackage{booktabs}
\usepackage{makecell}
\usepackage{multirow}  

\begin{document}

\preprint{APS/123-QED}

\title{Superradiance in acoustic black hole}

\author{Chengye Yu}
\email{chengyeyu1@hotmail.com}
\affiliation{School of Physics, Beijing Institute of Technology, Beijing, 100081, China}

\author{Xiaolin Zhang}
\email{xiaolinzhang66@outlook.com}
\affiliation{School of Physics, Beijing Institute of Technology, Beijing, 100081, China}

\author{Sobhan Kazempour}
\email{sobhan.kazempour1989@gmail.com}
\affiliation{School of Physics, Beijing Institute of Technology, Beijing, 100081, China}

\author{Sichun Sun}
\email{sichunssun@gmail.com}
\affiliation{School of Physics, Beijing Institute of Technology, Beijing, 100081, China}


\begin{abstract}
	Rotating superradiance in cylindrical geometries has recently been observed experimentally using acoustic waves, shedding light on the superradiant phenomenon in black holes.  
	In this paper, we study superradiance in acoustic black holes made with solid material for the first time, using theoretical analysis and numerical simulations in COMSOL Multiphysics. We find that superradiance can occur in acoustic black holes when the general superradiance condition is met. We also find that the amplification effect is significantly weaker in acoustic black holes than in regular cylinders, due to absorption within the black holes. Furthermore, we have found that different acoustic black hole models exhibit similar superradiance behavior at the same physical scale, which is also consistent with the phenomena in extremal Kerr black holes. We also present that the solid material ABH model has the most degrees of freedom.
\end{abstract}


\maketitle


\section{Introduction}\label{sec:1}

Dicke \cite{PhysRev.93.99} proposed the concept of superradiant states \cite{2015LNP...906.....B}, attributing its amplification to coherence. In 1969, Roger Pense proposed the energy extraction of a rotational black hole as the early version of the superradiance\cite {1969NCimR...1..252P,2002PER}. Later, Zel'Dovich \cite{1972JETP...35.1085Z,1971JETPL..14..180Z} suggests that superradiance can occur on the surface of rotating objects, where the amplitude of the incident wave is significantly larger than the amplitude of the reflected wave. (The phenomenon of superradiance in black holes has also been extensively studied; see \cite{2016PhLB..758..181H,2010PhRvD..81l3530A,2011PhRvD..83d4026A,2000PhRvL..84.4537A,2004PhRvD..70h4011C,2013PhRvD..87d3513W,2021JCAP...07..033M,2017PhRvD..96c5019B,PhysRevD.88.063003,2018PhLB..781..651D,2015CQGra..32m4001B,2019PhRvD..99j4030G,2008JHEP...06..071C,2018PhRvL.120w1102R,2017PhRvL.119d1101E,2016JCAP...12..043K,2015PhRvD..91l4018R,2016PhRvD..93f4066W,2014PTEP.2014d3E02Y,2020JHEP...08..105Z,2021PhRvD.104j3009S,2021LRR....24....4A}.) Related experiments are then proposed to detect this phenomenon, especially for rotational superradiance \cite{2015LNP...906.....B,2017JHEP...05..052E,2016PhRvL.117A1101C,Starobinsky:1973aij,1998PhRvD..58f4014B,PhysRevLett.123.044301,2017TTP,Andersson:1998swa}. In experiments, rotational superradiance occurs when an object moves around a rotational axis with a specific angular velocity, and the condition for superradiance is
\begin{eqnarray}
\omega -m\Omega <0,
\label{eq. 01}
\end{eqnarray}

\noindent where $\omega $ is the incident waves frequency, $m$, is the azimuthal quantum number with respect to the axis of rotation, and $\Omega $ is the angular frequency of the cylinder. Eq. (\ref{eq. 01}) represents the most general condition for superradiance, and it serves as an indicator for determining whether the phenomenon of superradiance occurs \cite{1945gvl,Cardoso:2005vk,2023PhRvL.131k1601S,2021NaPho..15..272Y,2022NaPho..16..707K,2024arXiv240809486J}. 

It is found that low-frequency sound modes with orbital angular momentum (OAM) \cite{1974RoyalSociety,PhysRevLett.100.024302} are transmitted through an absorbing rotating disk and amplified when the disk rotation rate meets superradiant conditions \cite{2020NatPh..16.1069C,2018PNAS}. This rotational Doppler effect \cite{PhysRevLett.80.3217,Skeldon_2008} has also been observed in light-carrying OAM that is backscattered from a rotating rough surface, enabling the remote measurement of an object’s rotation speed \cite{2013SCIENCE,20131002ScientificReports,PhysRevA.90.011801}. There are other methods to detect superradiance; for example, Teukolsky et al \cite{1974ApJ...193..443T} proposed the idea of creating a "perfect mirror" in a so-called black hole bomb\cite{1972PRE,Cardoso:2004nk,2012PhRvL.109m1102P}. Numerical studies of charged black hole bombs at the linear level in the field amplitude have been conducted in both the frequency and time domains, as in Refs. \cite{PhysRevD.88.063003,2014PhRvD..89f3005D,2018CQGra..35r4001D}, while analytical studies have been carried out in Refs. \cite{2013PhRvD..88f4055H,2015CoTPh..63..569L}.

Despite significant progress in the study of superradiance, there remains a lack of understanding of the phenomenon in certain systems, such as the acoustic ‘black hole’ (ABH). ABHs are systems that mimic the behavior of black holes in general relativity, but in the context of acoustic waves \cite{2015PhRvD..91l4018R,MattVisser1998,Berti:2004ju,Cardoso:2004fi,Lepe:2004kv}. The concept of ABH is derived from the gravitational effects of astronomical black holes \cite{1988MIR}. 
As is widely known, the information that approaches a black hole is absorbed and debatably transmitted outside the black hole in the form of Hawking radiation \cite{1974HAW,1976HAWW,PhysRevD.14.2460,2000PhRvL..85.5042P,2001PhRvD..64d4006H}. In engineering, researchers have discovered that altering the surface structure of a solid can cause an elastic wave propagating within it to be absorbed by a wedge structure, producing an effect analogous to the gravitational pull of a black hole. This phenomenon is referred to as ABH structure \cite{1989KVV,KRYLOV2004605,DENIS20142475,TANG2017116,PELAT2020115316}. The study of superradiance in ABHs is important because it can provide insights into black-hole behavior and the radiation emitted by rotating systems.

The paper is organized as follows: In Section \ref{a}, we introduce the ABH and derive its amplification factor by solving the radial equation subject to the relevant boundary conditions. 
In Section \ref{b}, we use semi-analytical and simulation methods in COMSOL Multiphysics \cite{Cheli2015, Zienkiewicz2005TheFE,2024comsol} to validate superradiance conditions and amplification factors. In Section \ref{c}, we summarize our main results and discuss the implications of our findings. Finally, in Section \ref{d}, we establish an analogy between the solid material ABH model and the rotating draining bathtub ABH model, and investigate the resulting superradiance.

\section{Acoustic Black Hole set-up}\label{a}

The physical structure of two-dimensional ABH is given by \cite{KRYLOV2004605}. The plate consists of a section with a constant thickness $h_{2}$ extending from $R$ to the edge of the plate, a tapered region (from $r_{1}$ to $R$), and a central plateau from 0 to $r_{1}$ with a constant thickness $h_{1}$. The shape is as follows:

\begin{eqnarray}
h(r)=\left\{\begin{matrix}
h_{2}\qquad \quad (r\ge R),
\\b(r-r_{1})^{n} +h_{1}\quad(r\le R).
\end{matrix}\right. 
\label{eq. 02}
\end{eqnarray}

\begin{figure*}
	\centering 
	\includegraphics[width=1\textwidth]{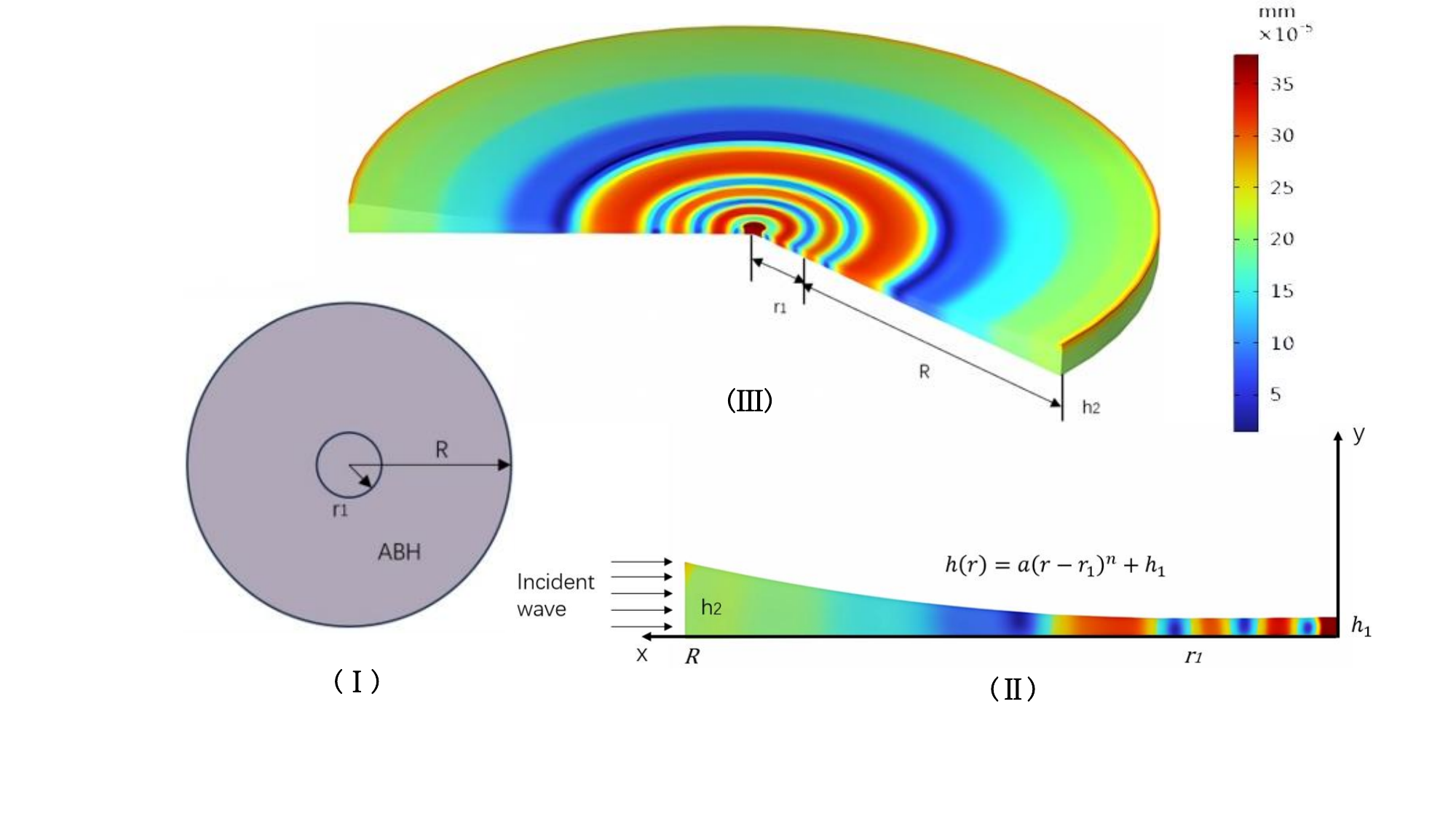} 
	\caption{(\uppercase\expandafter{\romannumeral1}) Sketch of a top view of ABH. (\uppercase\expandafter{\romannumeral2})A slice view of two-dimensional ABH simulated by COMSOL. (\uppercase\expandafter{\romannumeral3})3D view of ABH. Color scheme: Range of displacement induced by a $1 Pa$ point load applied at the boundary of the ABH, where the simulation is transient and the azimuthal mode number $m=0$.} 
	\label{fig. 0} 
\end{figure*}

The extreme case with $h_{1}=r_{1}=0$ and $n\ge 2$ corresponds to an ideal ABH structure (see Fig. \ref{fig. 0}). In this paper, we set the following parameters for ABH structure: $b=7.34\times 10^{-4}m^{-1}$, $r_{1}=2\times 10^{-2} m$, $h_{1}=6\times 10^{-4} m$ and $n=2$ (correspond to the \textbf{Simulation analysis} in Sec. \ref{c}), and the superradiance analytical analysis in ABH is derived in Sec. \ref{b}.

\section{Superradiance Mode}\label{b}

We start from Kirchhoff’s equation \cite{10.1121/1.399390} for a quiescent inviscid and non-conducting fluid, with the equation of mass conservation $\partial \rho /\partial t+\nabla \rho\bm{\nu }=0$ and the momentum conservation $\rho\partial \bm{\nu }/\partial t=-\nabla p$. In the air $(r>R)$ we have
\begin{eqnarray}
&&i\omega \rho_{+}+\rho _{0}\nabla \bm{\nu _{+}}=0,\nonumber\\
&&i\omega \rho_{0}\bm{\nu _{+}}+\nabla p _{+}=0,\nonumber\\
&&p_{+}=c_{s}^{2}\rho _{+},
\label{eq. 001}
\end{eqnarray}

\noindent while the ABH is described by the model $(r<R)$
\begin{eqnarray}
&&i\omega \rho_{-}+\rho _{f}\nabla \bm{\nu _{-}}=0,\nonumber\\
&&i\omega \rho_{f}\bm{\nu _{-}}+\nabla p _{-}=0,\nonumber\\
&&p_{-}=c^{2}_{eff}\rho _{-},
\label{eq. 002}
\end{eqnarray}

\noindent where $R$ is the radius of ABH. Zel’dovich’s original “dynamical” argument \cite{1972JETP...35.1085Z,1971JETPL..14..180Z} follows the Lorentz-invariant Klein-Gordon equation, and through Eq. (\ref{eq. 001}) , we have $\Box p_{+}=0$ outside the ABH.

In a coordinate system where the medium is at rest, the absorption can be characterized by a parameter $\alpha $, causing Eq. (\ref{eq. 002}) to become a modified equation 
\begin{eqnarray}
\Box p_{-}=\frac{\alpha }{c^{2}_{eff}}\partial _{t}p_{-}
\label{eq. 003}
\end{eqnarray}


We reconsider the wave equation (\ref{eq. 001}) outside the ABH in cylindrical coordinates, and the radial equation reduces exactly to 
\begin{eqnarray}
\frac{1}{r}\frac{\partial }{\partial r}(r\frac{\partial p_{+}(r)}{\partial r})+[\frac{\omega ^{2}}{c_{s}^{2}}-\frac{m^{2}}{r^{2}}]p_{+}(r)=0.
\label{eq. 1}
\end{eqnarray}

Dissipation is essential for the generation of the superradiance. In the rotating coordinate system, $\alpha (\omega -m\Omega )$ replaces $\alpha \omega$ under the superradiation condition, meaning that the absorption term becomes an amplification term, allowing superradiance to occur \cite{2015LNP...906.....B}. Inside the ABH, the wave equation (\ref{eq. 003}) can be rewritten in the more convenient form 
\begin{eqnarray}
\frac{1}{r}\frac{\partial }{\partial r}(r\frac{\partial p_{-}(r)}{\partial r})+[\frac{\kappa ^{2}}{c_{eff}^{2}(r)} -\frac{m^{2}}{r^{2}}]p_{-}(r)=0,
\label{eq. 2}
\end{eqnarray}

\noindent where
\begin{eqnarray}
\kappa =\sqrt{\omega ^{2}+i\alpha (\omega -m\Omega)}, \quad c_{eff}(r)=\sqrt{c_{s}^{2}-v^{2}(r)},
\label{eq. 3}
\end{eqnarray}

\noindent where $\kappa $ is the complex angular frequency. The ABH contains an effective complex sound velocity $c_{eff}(r)$ (where $c_{s}$ is the sound speed) relative to the radial distance $r$, with $v(r)$ representing the background flow rate\cite{MattVisser1998}. The solution to Eq. (\ref{eq. 1}) is the combination of Bessel functions, 
\begin{eqnarray}
p_{+}(r)=C_{1}J_{m}(\omega r/c_{s})+C_{2}Y_{m}(\omega r/c_{s}), 
\label{eq. 4}
\end{eqnarray}

\noindent where the coefficients $C_{1}$ and $C_{2}$ are related to the incident (reflected) part with amplitude for outside ABH. It is worth noting that $\omega /c_{s}$ is the number of sound waves outside ABH. It is evident that the solution to Eq. (\ref{eq. 2}) depends on the background flow rate $v(r)$. Notably, whether the background flow rate is constant or not significantly influences the internal solutions of the ABH. Therefore, this aspect will be the focus of discussion in the next section.

Before working out explicitly the case of a rotating object in $(r,\varphi ,z)$ spatial coordinates with a dissipative surface at $r=R$, we need consider a small quantity $\varepsilon$ relative to $R$ that outside surface $R_{+}$ is $R+\varepsilon $ and inside surface $R_{-}$ is $R-\varepsilon $. 
\begin{figure}
	\centering 
	\includegraphics[width=0.48\textwidth]{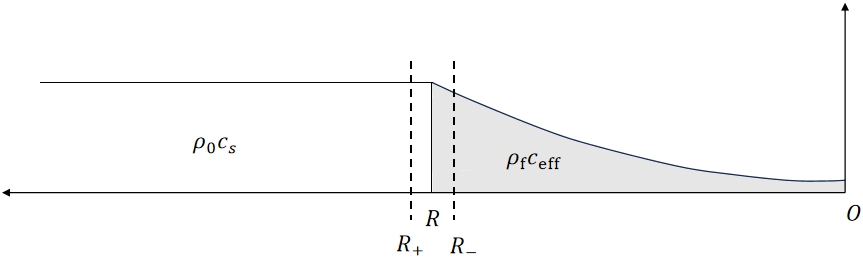} 
	\caption{Outside impedance and inside impedance of the ABH.} 
	\label{fig. 01} 
\end{figure}
Obviously, in Fig. \ref{fig. 01}, a discontinuity exists between the outer surface $R_{+}$ and the inner surface $R_{-}$, requiring continuity conditions to ensure smoothness on the scattering surface. We consider matching the pressure and velocity of the internal and external solutions under the jump condition of $r=R$\cite{2021SWRA}. From the continuity of velocity $(\bm{\nu_{+}}=\bm{\nu_{-}})$ in Eqs. (\ref{eq. 001}), (\ref{eq. 002}), we can get
\begin{eqnarray}
\frac{\partial p_{+}(r)}{\partial r}|_{r=R_{+}}=\frac{\rho _{0}}{\rho _{f}} \frac{\partial p_{-}(r)}{\partial r}|_{r=R_{-}}.
\label{eq. 9}
\end{eqnarray}

We can define the impedances inside and outside the ABH. Since sound scattering occurs at the boundary of the ABH, we assume the small quantity is negligible, allowing us to obtain the associated impedance as
\begin{eqnarray}
Z_{+}=\rho_{0}c_{s},\quad Z_{-}=\rho_{f}c_{eff}(r)|_{r=R_{-}}=\rho_{0}c_{s}Z,
\end{eqnarray}

\noindent where $Z_{+}$ is the outside impedence, $Z_{-}$ is the inside impedence, and $Z$ is the characteristic acoustic impedance of the fibrous material\cite{1970DEM} (see Fig. \ref{fig. 01}).

Due to the mismatch in acoustic impedance between the external and internal environments, the incident wave encounters inside impedance conditions at the scattering interface $r=R$. In terms of impedance, the boundary condition for both sound and surface waves in the rest frame of ABH is
\begin{eqnarray}
p_{+}(R_{+})-p_{-}(R_{-})=-\frac{Z_{-}}{i\rho _{0}\omega }\frac{\partial p_{+}(r)}{\partial r}|_{r=R_{+}}.
\label{eq. 7}
\end{eqnarray}


When the object rotates uniformly with angular velocity $\Omega $, it is sufficient to transform to a new angular coordinate $\tilde{\phi }=\phi +\Omega t$, which effectively amounts to the replacement of $\omega $ with $\omega -m\Omega $ in Eq. (\ref{eq. 7}). Then the inside impedance condition (\ref{eq. 7}) becomes
\begin{eqnarray}
p_{+}(R_{+})-p_{-}(R_{-})=-\frac{Z_{-}}{i\rho _{0}(\omega -m\Omega )}\frac{\partial p_{+}(r)}{\partial r}|_{r=R_{+}}.
\label{eq. 10}
\end{eqnarray}

Then through Eq. (\ref{eq. 9}), (\ref{eq. 10}), the relationship between coefficients $C_{1}$ and $C_{2}$ is
\begin{eqnarray}
\frac{C_{1}}{C_{2}}=-\frac{\tilde{Z}{Y}'_{m}(\omega R_{+}/c_{s} )-Y_{m}(\omega R_{+}/c_{s})}{\tilde{Z}{J}'_{m}(\omega R_{+}/c_{s} )-J_{m}(\omega R_{+}/c_{s})}, 
\label{eq. 12}
\end{eqnarray}

\noindent where
\begin{eqnarray}
\tilde{Z}=Z(\frac{\omega }{c_{eff}(r)}\frac{p_{-}(r)}{\partial_{r}p_{-}(r)}|_{r=R_{-}}  -\frac{\omega }{i(\omega -m\Omega) } ) ,
\label{eq. 14}
\end{eqnarray}

\noindent where $'$ means derivative with respect to the argument. The relationship between the reflected (incident) part with amplitude and coefficients $C_{1}$, $C_{2}$ is $C_{1}-iC_{2}$ $(C_{1}+iC_{2})$. For simplicity, letting $C=C_{1}/C_{2}$, and amplification factor can be written as
\begin{eqnarray}
\rho =\left | \frac{C-i}{C+i} \right |^{2}-1. 
\label{eq. 13}
\end{eqnarray}

One of the conditions for rotational superradiance is $\omega -m\Omega$, where the superradiant phenomenon occurs on the radiating surface of the cylinder, that is, $\rho >0$. It is worth mentioning that if the cylinder is conventional and not ABH inside, then the amplification factor (\ref{eq. 13}) is different, which we will discuss more in the section \ref{c}.

\section{semi-analytical and Simulation Analysis}\label{c}

In this section, we perform semi-analytical and COMSOL simulations to study the superradiance relation of the sound wave angular frequency and cylinder/ABH angular velocity. The two methods obtain amplification factors for a cylinder and ABH, respectively, and agree well. First, we provide a brief overview of the analytical methods to ensure consistency between the theoretical and simulation analyses.

\textbf{Semi-analytical analysis:} We use Eq. (\ref{eq. 13}) to calculate amplification factor, for semi-analytical calculation

\textbf{Simulation analysis:} In this paper, COMSOL software is used to simulate the acoustic superradiant phenomenon occurring in a fluid domain, and the amplification factor $\rho _{s}$ of the cylinder or ABH is obtained by setting appropriate boundary conditions with the pressure acoustic module and Delany-Bazley-Miki model \cite{1970DEM,199019}. The amplification factor $\rho _{s}$ is the ratio between the background sound pressure level and the scattered sound pressure level.


\textbf{Settings:} The values of the parameter in Eq. (\ref{eq. 13}) are fixed as: the density of air is $1.2 kg/m^{3}$, the sound speed of air $c_{s}$ is $343 m/s$, the radius of the cylinder $R$ is $10^{-1} m$ and the azimuthal quantum number is $m=1$. The value of the sound absorption parameter $\alpha$ does not impact much on the superradiant intensity. For simplicity we set $\alpha =0s^{-1}$. Other parameters are mentioned in section \ref{a}.

When the background fluid rate $v(r)$ is constant, Eq. (\ref{eq. 2})  is just the inner solution of the cylinder, and its solution can also be expressed as the Bessel function
\begin{eqnarray}
p_{-}(r)=C_{3}J_{m}(\kappa r/c_{eff}),
\label{eq. 15}
\end{eqnarray}

\noindent When the background fluid rate changes, if we assume that the rate $v(r)$ changes gently with radial $r$, then approximated by WKB, the internal solution of the ABH is
\begin{eqnarray}
p_{-}(r)\sim \frac{1}{\sqrt[4]{rc_{eff}(r) }}exp(i\int \frac{\kappa }{c_{eff}(r)}dr ).
\label{eq. 16}
\end{eqnarray}

The characteristic acoustic impedance of the fiber material is
\begin{eqnarray}
Z=1+A_{1}X^{-A_{2}}-iA_{3}X^{-A_{4}}, X=\rho_{0}\omega/(2\pi \varpi ),
\label{eq. 17}
\end{eqnarray}

\noindent where $\varpi $ is the flow resistance of the material, $\rho_0$ is the air density. For the small frequency that  $X\leq0.025$, we can use the values below. We mainly examine glass fiber and rock materials, with their specific acoustic impedance parameters shown in Table \ref{tab. 0}.
\begin{table}[htbp]
	\centering
	\renewcommand{\arraystretch}{1.5} 
	\setlength{\tabcolsep}{3pt} 
	\caption{Characteristic acoustic impedance for Miki and low-frequency Glass (Rock) fiber.}
	\begin{tabular}{cccccc}
		\hline
		Materials\cite{Mechel2004FormulasOA}&$X$&$A_{1}$&$A_{2}$&$A_{3}$&$A_{4}$\\
		\hline
		Miki&$<1$&0.079&0.632&0.12&0.632\\
		Low Glass fibre(LG) &$\le 0.025$&$0.0688$&$0.707$&$0.196$&$0.549$\\
		Low Rock fibre(LR) &$\le 0.025$&$0.081$&$0.699$&$0.191$&$0.556$\\
		\hline
	\end{tabular}
	\label{tab. 0}
\end{table}
\begin{table}[htbp]
	\centering
	\renewcommand{\arraystretch}{1.5} 
	\setlength{\tabcolsep}{5pt} 
	\caption{The different values of flow resistances we used in the plot.}
	\begin{tabular}{ccccc}
		\hline
		Flow resistance ($\varpi $)&$\varpi _{0}$&$\varpi _{1}$&$\varpi _{2}$&$\varpi _{3}$\\
		\hline
		Value ($Pa\cdot s/m^{2}$)&$2.5*10^{3}$&$1*10^{4}$&$2*10^{4}$&$5*10^{4}$\\
		\hline
	\end{tabular}
	\label{tab. 1}
\end{table}

It is worth mentioning that two conditions are essential: the first is the impedance of the fibrous material attenuation term $Z$, and the second is the superradiant condition $\omega -m\Omega <0$, which determines whether the superradiance can occur. 

\subsection{Simulation of acoustic superradiation in a fluid domain}

\subsubsection{Sound field ABH in the non-rotational case}

This section investigates the sound field distribution of ABHs in a fluid domain. First, we analyze the sound-field distribution in a static fluid domain, before extending the study to a rotating fluid domain.

\begin{figure}
	\centering 
	\includegraphics[width=0.48\textwidth]{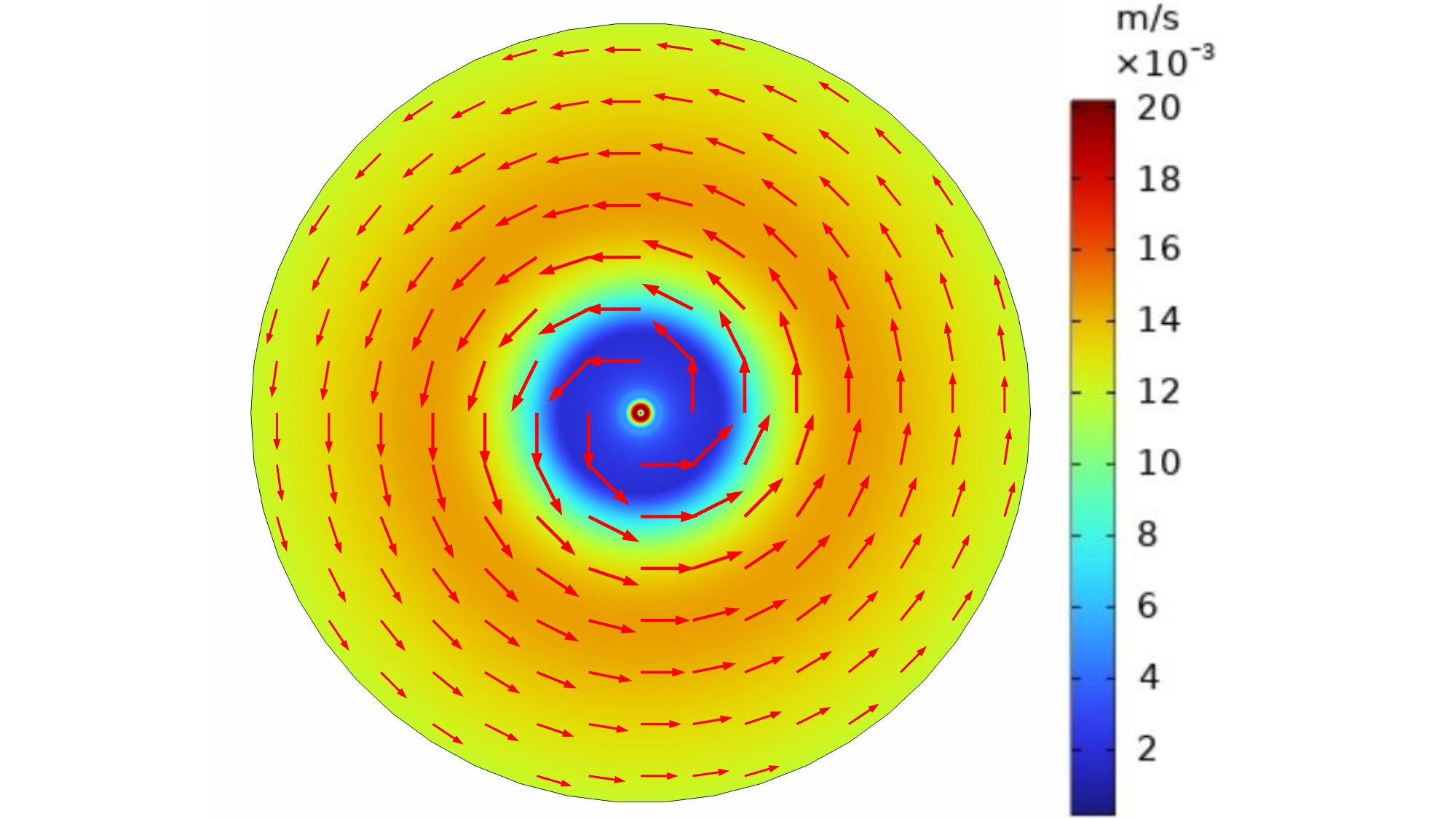} 
	\caption{The distribution of a sound field in an ABH where the frequency of the sound wave is $100 Hz$ and the angular velocity of the ABH is $0 rad/s$.
		\textit{Color scheme}: Range of transient sound speed induced by the specified acoustic pressure field } 
	\label{fig. 1-1} 
\end{figure}

Fig. \ref{fig. 1-1} illustrates the distribution of sound waves with an azimuth mode $m=1$ within an ABH in a quiescent flow field. The fluid flow field comprises both angular and radial velocity components, characterized by the azimuthal number m. The effect of the acoustic black hole is that the sound speed decreases towards the singularity. Notice the artifact we put in by hand near the center plateau region in this simulation.

\subsubsection{Rotational Superradiance}

The key to acoustic superradiation is the superradiation condition, where the relation of the angular frequency $\omega $ and rotational frequency $\Omega$ enables superradiation amplification. 
In this section, the angular velocity of the acoustic black hole is set to be $200\pi rad/s$ with different incident sound wave frequencies. 

\begin{figure}
	\centering 
	\includegraphics[width=0.48\textwidth]{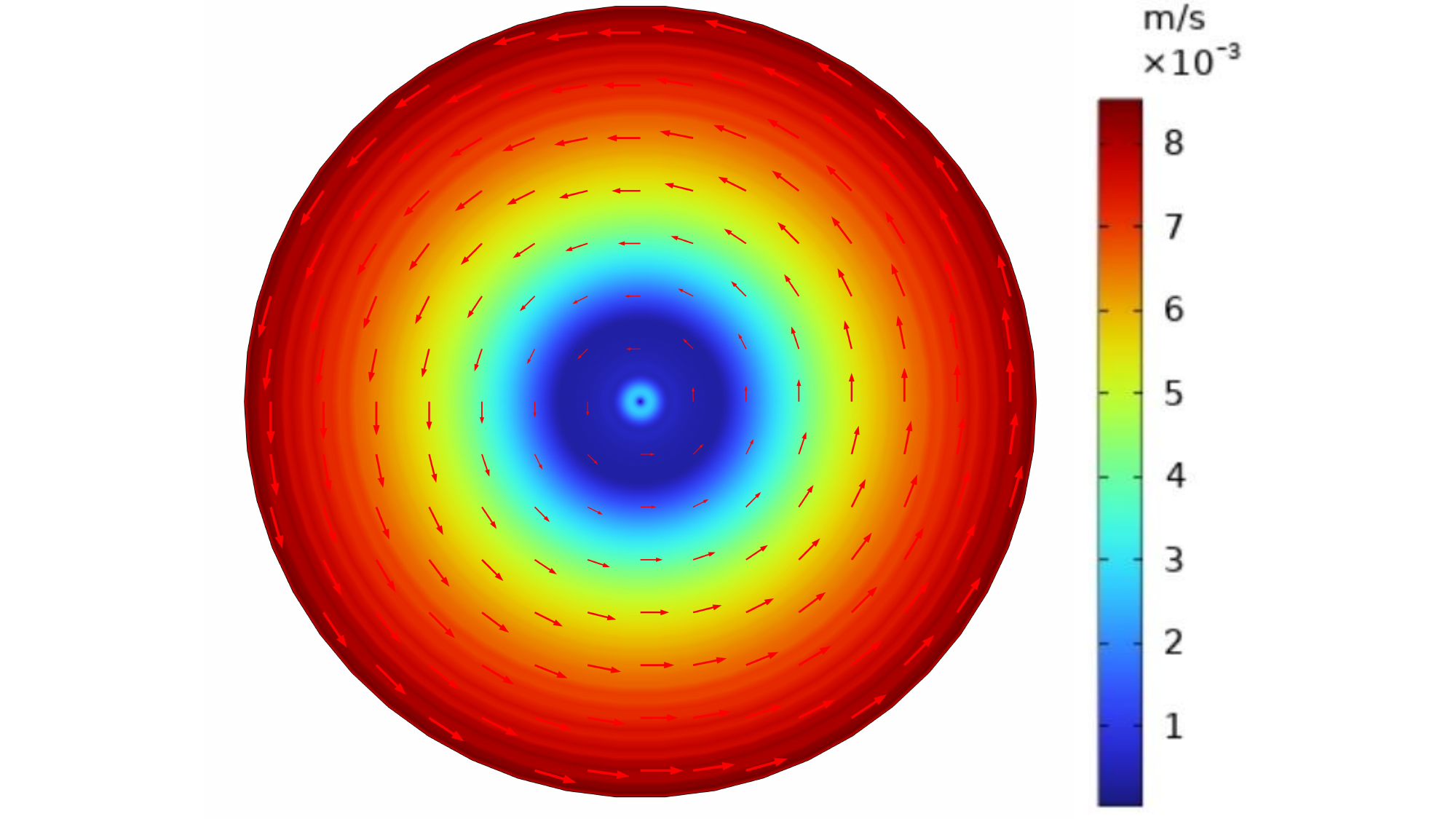} 
	\caption{The distribution of a sound field in an ABH where the frequency of the sound wave is $200 Hz$, and the angular velocity of the ABH is $200\pi rad/s$. Color scheme: range of transient sound speed induced by the specified acoustic pressure and the rotation.}
	\label{fig. 1-2}  
\end{figure}

Fig. \ref{fig. 1-2} illustrates the non-superradiation case. In this case, the sound waves tend to converge at the black hole’s ‘event horizon’ rather than its ‘singularity’, dividing the sound field into three regions: the outer high-velocity region (near the ‘event horizon’), the middle medium-velocity region, and the inner low-velocity region (near the ‘singularity’). We can see the radial rotational effect in this plot compared to the static Fig.\ref{fig. 1-1}.

\begin{figure}
	\centering 
	\includegraphics[width=0.48\textwidth]{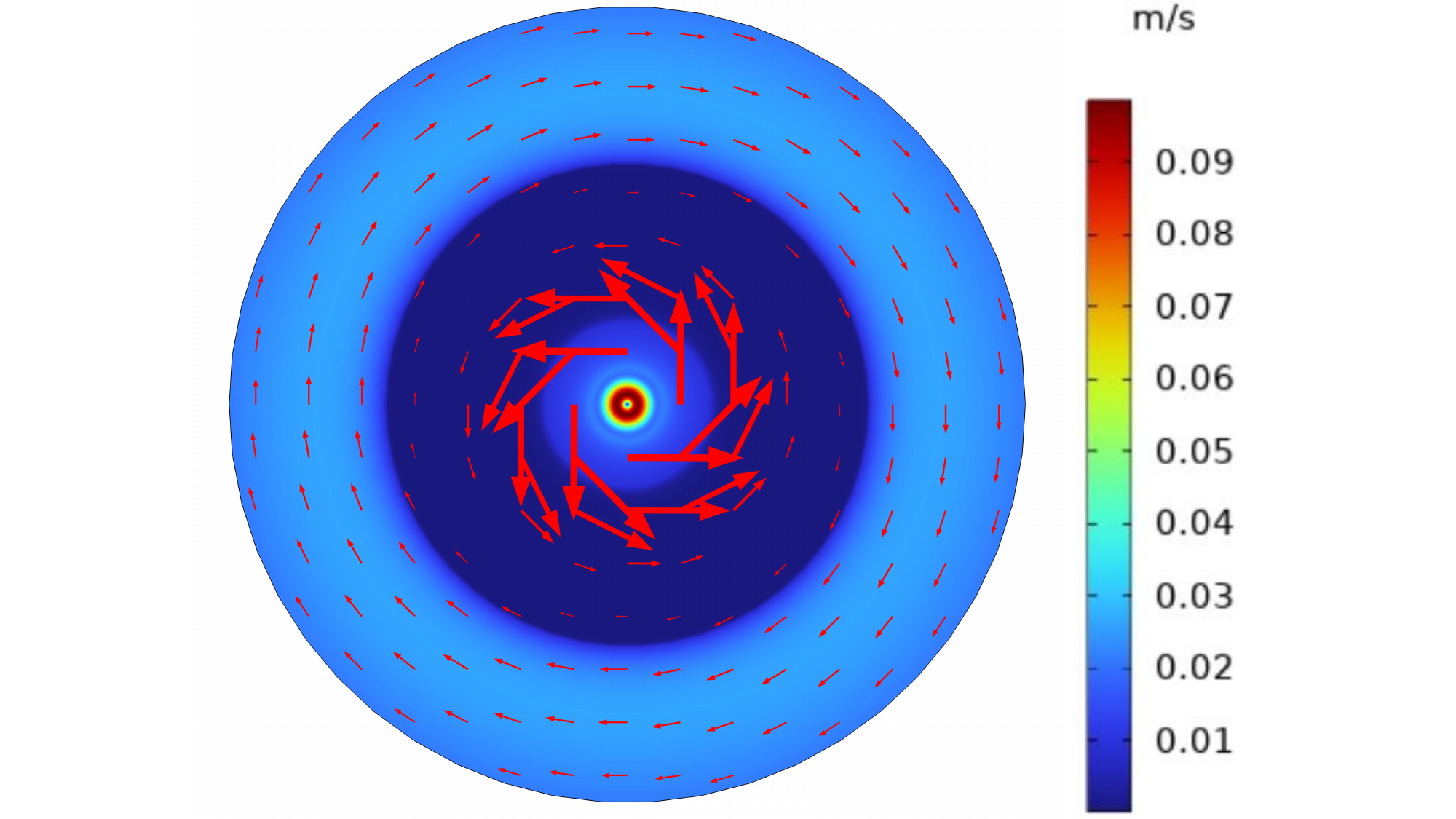}
	\caption{The distribution of a sound field in an ABH where the frequency of the sound wave is $20 Hz$, and the angular velocity of the ABH is $200\pi rad/s$. The propagation direction of sound waves is in the radial direction of the ABH. Color scheme: Range of transient sound speed induced by the specified acoustic pressure field and the rotation.}
	\label{fig. 1-3}  
\end{figure}

Fig. \ref{fig. 1-3} satisfies the superradiation condition, and it has instability in the fluid domain. In the radial direction, the sound wave propagates outward in some region, effectively indicating that the superradiation condition $\omega -m\Omega$ replaces the inherent angular frequency $\omega$ of the sound wave. 
This observation is critical in detecting the amplification effect of acoustic superradiation and provides a foundation for future experimental observations of superradiation phenomena.

\subsection{Superradiance for ABH}

Unlike the amplification factor $\rho $ of the semi-analytical analysis, the amplification factor of the COMSOL simulation can be written as
\begin{eqnarray}
L_{p,s}-L_{p,b}&&=20log_{10}(\frac{p_{s}}{p_{ref}})-20log_{10}(\frac{p_{b}}{p_{ref}})\nonumber\\
&&=20log_{10}(R_{s,b})
\label{eq. 18}
\end{eqnarray}

\noindent where $L_{p,s}$ is the scattering sound pressure level and $L_{p,b}$ is the background sound pressure level. $R_{s,b}$ represents the reflection coefficient, and the reference sound pressure $p_{ref}=20\mu Pa$. In the simulation, with an incident wave amplitude of $1 Pa$ in air, the background sound pressure level $L_{p,b}$ is measured at $ 90.969 dB$. 

\begin{figure}
	\centering 
	\includegraphics[width=0.48\textwidth]{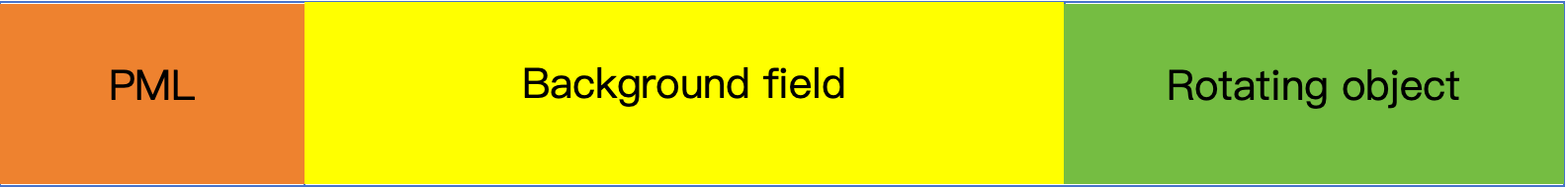}
	\caption{The main components of the simulation. PML: Absorbs the outgoing waves and monitors the intensity of the scattered waves.  Background field: Provides the incident wave and is the region where superradiance occurs. Rotating object: Superradiance object.}
	\label{fig. 1-4}  
\end{figure}

\begin{figure*}
	\centering 
	\includegraphics[width=0.48\textwidth]{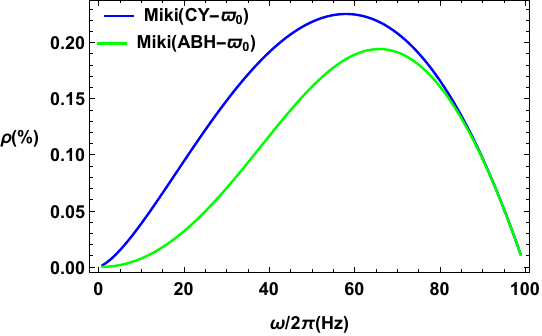}
	\includegraphics[width=0.48\textwidth]{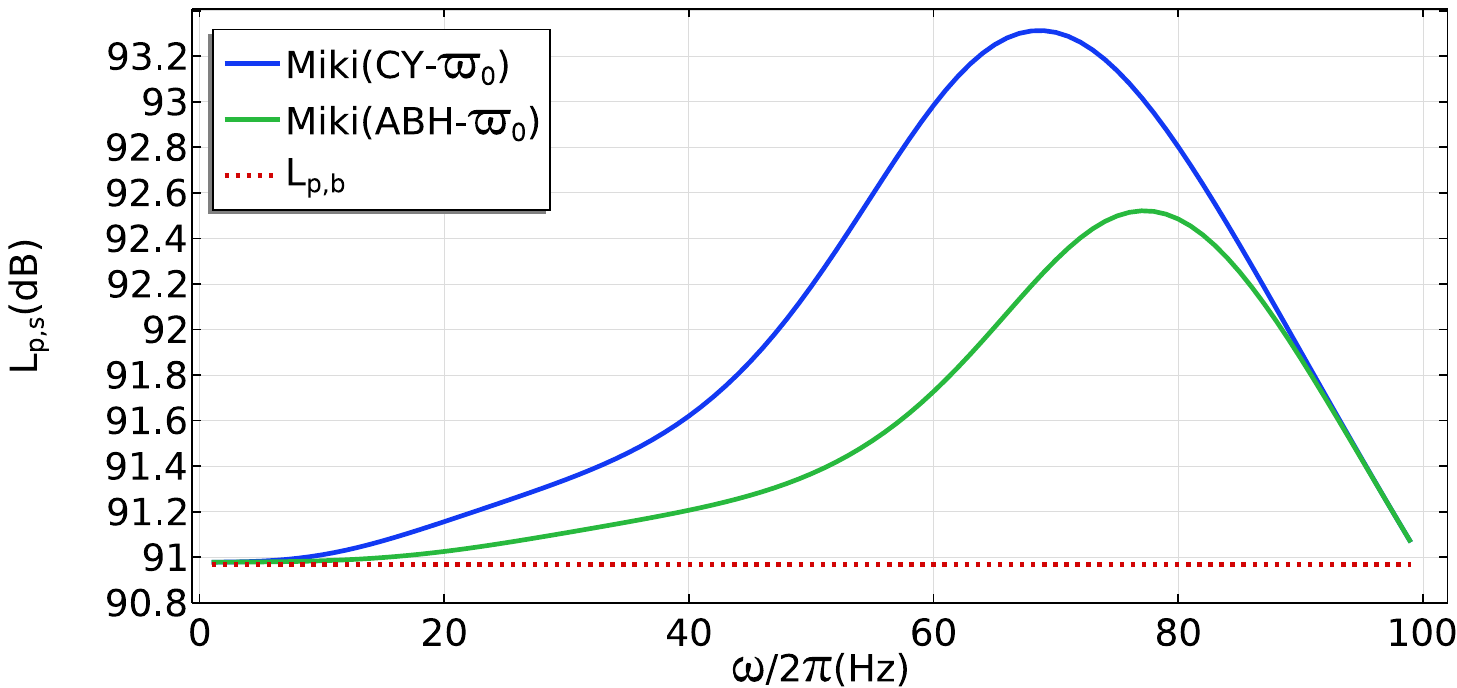}
	\includegraphics[width=0.48\textwidth]{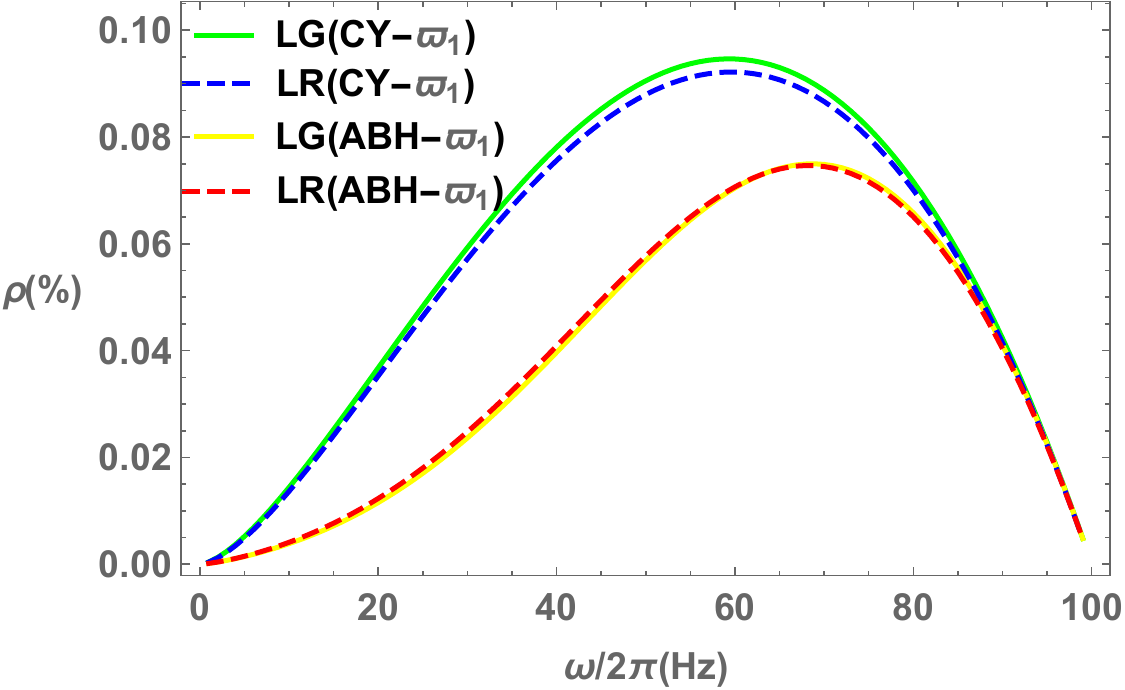}
	\includegraphics[width=0.48\textwidth]{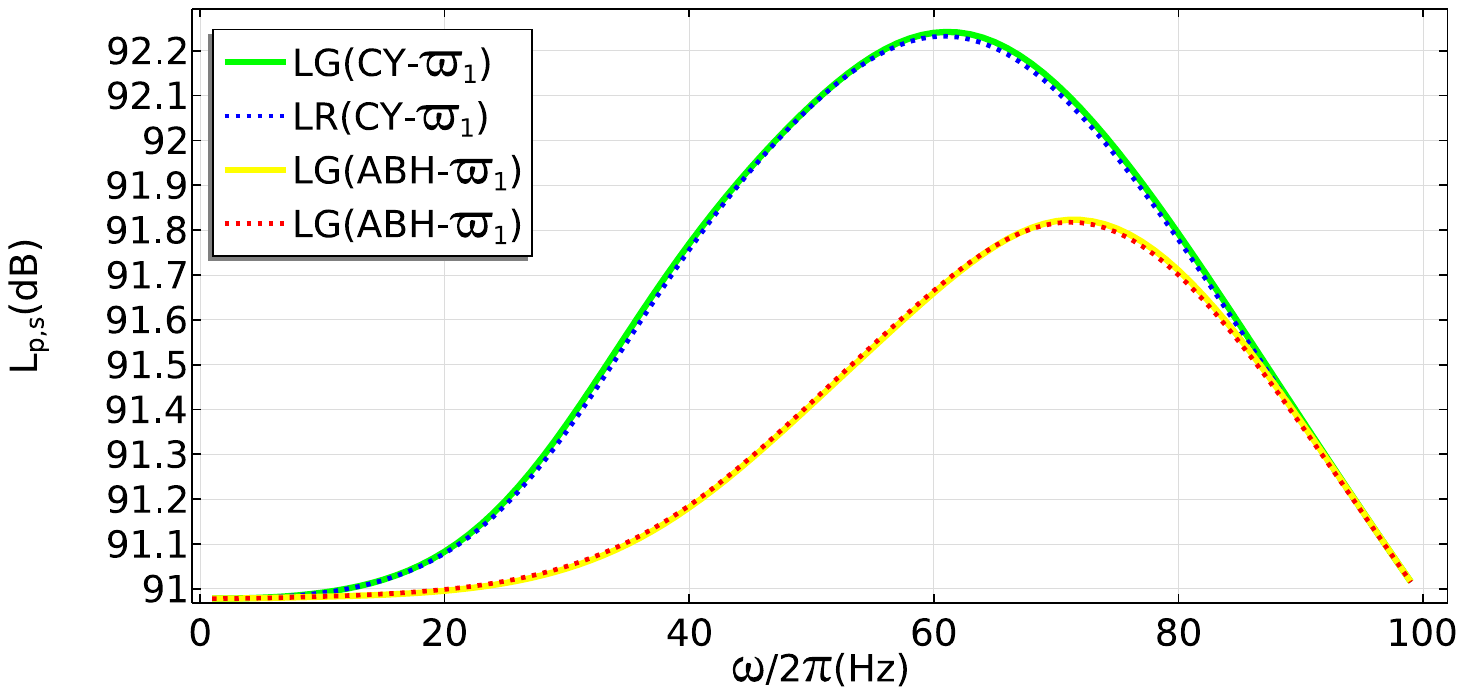}
	\includegraphics[width=0.48\textwidth]{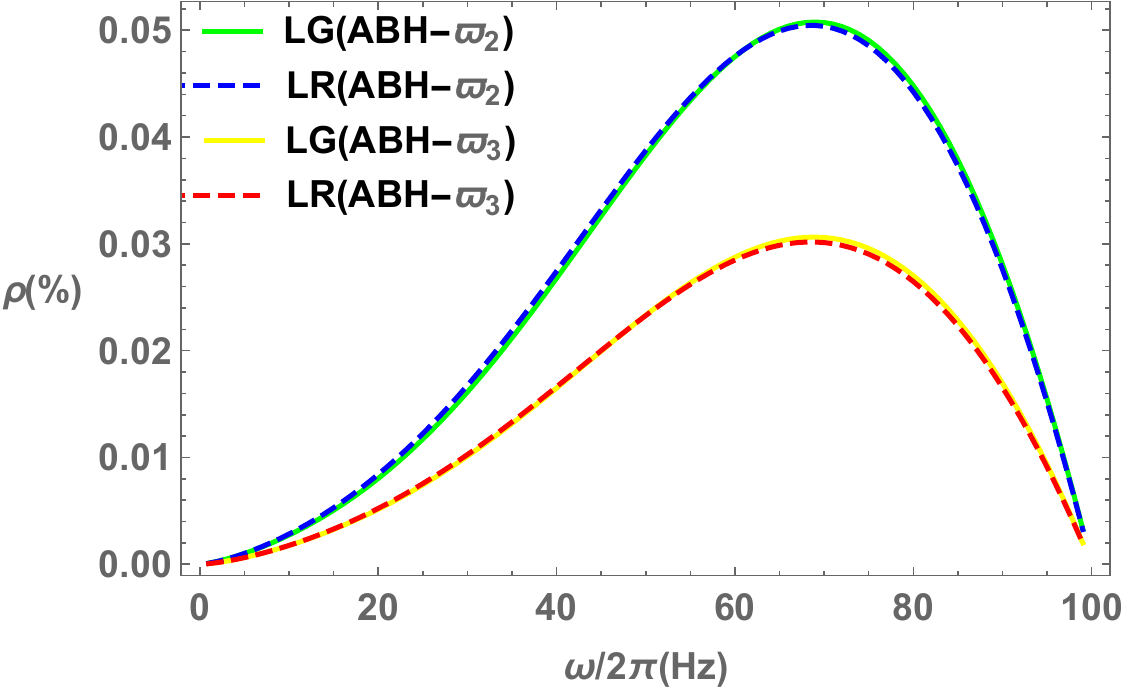}
	\includegraphics[width=0.48\textwidth]{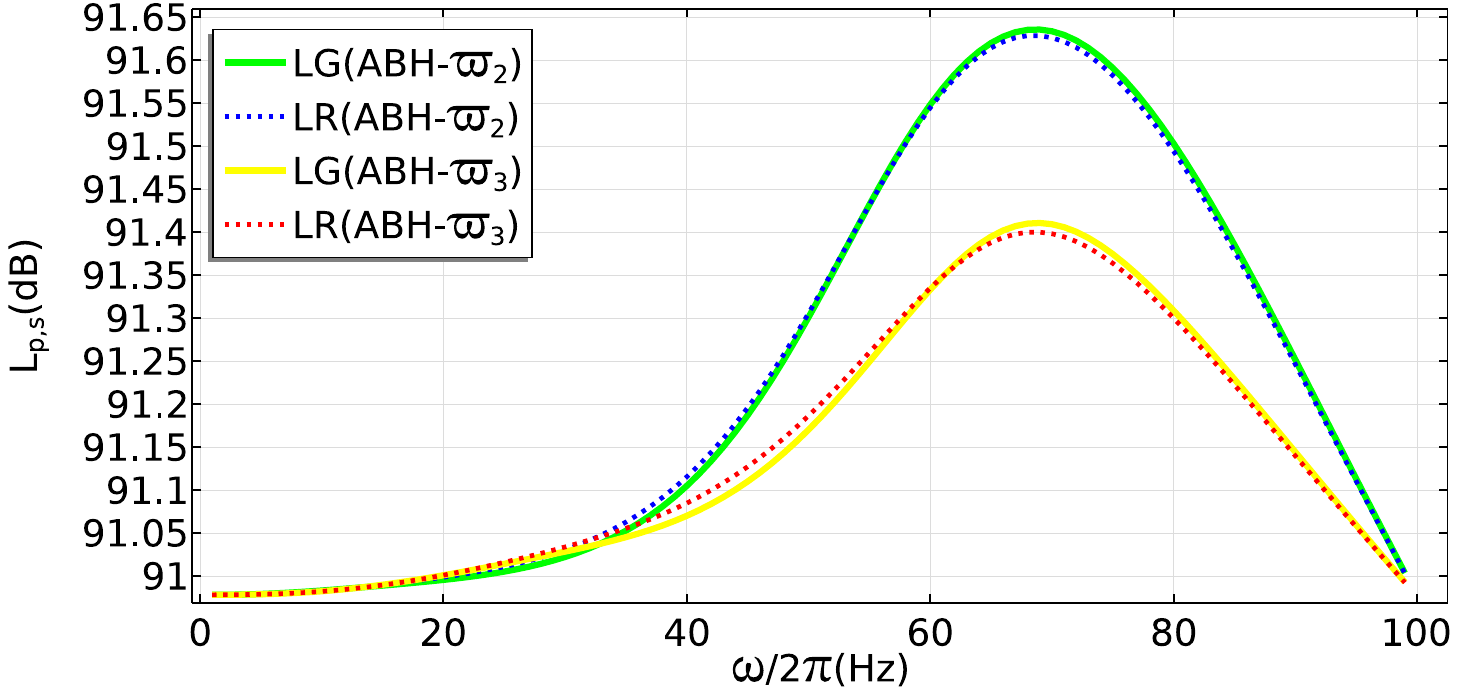} 
	\caption{The left figures show the amplification factor $\rho $ under semi-analytical analysis, and the right figures show the pressure level $\text{L}_{p,s}$ under simulation. CY stands for cylindrical structure, and ABH stands for ABH structure. }
	\label{fig. 2}
\end{figure*}

The main components of the simulation and their respective functions are depicted in Fig. \ref{fig. 1-4}. Absorption and superradiant amplification in an ABH are competing mechanisms in both semi-analytical and simulation.  In Fig. \ref{fig. 2}, we can see the different amplification factors (defined differently) obtained from the theoretical semi-analytical study and simulation.  The Delany-Bazley-Miki model was used to describe the acoustic impedance of fiber materials. This model better captures the frequency-dependent dissipation in real materials, resulting in stronger absorption. The actual acoustic impedance of the surface fiber layer failed to reach the theoretically assumed optimal value that maximizes absorption. Theoretically, in an ideal ABH, virtually all the incident wave energy is absorbed, with only a negligible portion being reflected. However, in practical implementation/simulation, the ABH tip cannot be infinitely sharp and is inevitably truncated, resulting in a small flat plateau Eq.\ref{eq. 02}, which substantially diminishes the efficiency of energy trapping and dissipation. More incident wave energy is reflected rather than absorbed. When this part of the reflected wave meets the hyperradiative condition at the rotational boundary, it can extract energy from the rotation of the system. Ultimately, the observed superradiation amplification factor appears to far exceed the semi-analytical factor in numerics in Fig \ref{fig. 2}.
 
Fig. \ref{fig. 2} shows that when the impedance $Z$ changes with increasing flow resistance of the fiber material, the amplification factor decreases. This suggests that the fiber's material properties significantly affect the superradiant phenomenon. Furthermore, ABH exhibits strong absorption due to its gradient thickness profile $h(r)$ and viscoelastic material losses. This absorption suppresses wave magnification in the ABH to approximately two-thirds of the value observed in a conventional cylinder. It also suggests that when the flow resistance of the fiber material becomes sufficiently large, the observation of superradiance can be considerably challenging.

Unlike a regular cylinder, ABH exhibits a strong absorption effect, causing the sound waves to be largely trapped within ABH. This phenomenon is clearly illustrated in Fig. \ref{fig. 2}, and the ABH internal structure causes the absorption of sound waves in the simulation. For a regular cylinder, the amplification factors are similar to the results of V. Cardoso et al\cite{2015LNP...906.....B,2016PhRvL.117A1101C}.

\section{Superradiance for analogue black hole}\label{d}

In this section, we study a different analogue black hole model: a draining bathtub fluid flow with a sink at the origin, firstly introduced by Visser \cite{MattVisser1998}. The superradiance of this kind of analogue black hole has been studied in previous works \cite{Berti:2004ju,Lepe:2004kv}. Interestingly, Ref. \cite{2015PhRvD..91l4018R} considered varying water depth and describes a more realistic draining geometry. 
In our Chapter \ref{e}, the rotating draining bathtub model is studied with the same parameter range as the solid material ABH, and the superradiance condition was derived. Chapter \ref{f} employs Eq. \ref{eq. 26} to compare the superradiance intensity 
across an identical parameter range. In Chapter \ref{g}, we present Tab. \ref{tab.4} to make analogies and summaries of the superradiance theoretical models:  the solid material  ABH, the draining bathtub, and the Kerr black hole.

At last, we conclude that across these three models, although the superradiance behavior is similar,  the solid material ABH has the most degrees of freedom.

\subsection{Rotating draining bathtub ABH}\label{e}

In this section, we review the basic setup of the draining bathtub model \cite {2015PhRvD..91l4018R}. The event horizon is defined as the surface where the radial velocity equals the propagation speed
\begin{eqnarray}
\bar{c}_{\text{gw}}^{2}=\bar{v}_{r}^{2}.
\label{eq. 19}
\end{eqnarray}

Here, we further normalize the gravitational wave velocity and the radial velocity as $\bar{c}_{\text{gw}}=c_{\text{gw}}/(\sqrt{g\textbf{h}_{\infty}})$ and $\bar{v}_{r}=v_{r}/(\sqrt{g\textbf{h}_{\infty}})$, so that we can directly obtain the dimensionless event horizon. $g$ is the gravitational acceleration and $\textbf{h}_{\infty}$ is the water depth. The wave equation around our analogue black hole can be written as
\begin{eqnarray}
\frac{d^2H(\bar{r}_{*})}{d\bar{r}_{*}^2}+\bar{V}(\bar{r})H(\bar{r}_{*})=0, 
\end{eqnarray}

\noindent where
\begin{eqnarray}
\bar{V}(\bar{r})&=&\frac{1}{\bar{g}\bar{\textbf{h}}}(\sigma \bar{\omega}-\sqrt{2}\frac{m\bar{B}}{\bar{r}^2})^2+\frac{1}{2}(\frac{(\bar{\textbf{h}}\bar{r})_{,\bar{r} }}{\bar{\textbf{h}}\bar{r}} )(\frac{\Delta_{,\bar{r} }}{\Delta^3}) \nonumber\\
&-&\frac{m}{\Delta\bar{r}^{2} } -\frac{1}{\Delta^2} [\frac{1}{4}(\frac{(\bar{\textbf{h}}\bar{r})_{,\bar{r} }}{\bar{\textbf{h}}\bar{r}} )^2+\frac{1}{2}(\frac{(\bar{\textbf{h}}\bar{r})_{,\bar{r} }}{\bar{\textbf{h}}\bar{r}} )_{,\bar{r} } ],
\label{eq. 20}
\end{eqnarray}

\noindent and its solution is
\begin{eqnarray}
H\sim \left\{\begin{matrix}
exp(-i\sigma \bar{\omega} \bar{r}_{*})+\mathcal{R} exp(i\sigma \bar{\omega}\bar{r}_{*}),\quad  \bar{r}\to \infty ,\\\mathcal{T}exp(-i\frac{1}{\sqrt{\bar{g}_{H}\bar{\textbf{h}}}_{H}}(\sigma\bar{\omega}-\frac{\sqrt{2}m\bar{B}}{\bar{r}_{H}^2})\bar{r}_{*} ), \bar{r} \to \bar{r}_{H} .
\end{matrix}\right.
\label{eq. 21}
\end{eqnarray}

Using boundary conditions (\ref{eq. 21}), we find the
following “energy conservation” condition
\begin{eqnarray}
1-|\mathcal{R}|^{2}=\frac{1}{\sqrt{\bar{g}_{H}\bar{\textbf{h}}}_{H}}(1-\frac{\sqrt{2}m\bar{B}}{\sigma \bar{\omega} \bar{r}_{H}^2})|\mathcal{T}|^{2},
\label{eq. 22}
\end{eqnarray}

\noindent and superradiance occurs whenever
\begin{eqnarray}
0 < \sigma \bar{\omega} < \frac{\sqrt{2}m\bar{B}}{\bar{r}_{H}^2},
\label{eq. 23}
\end{eqnarray}

\noindent where $\bar{\omega }=\omega /(\sigma \sqrt{g/\textbf{h}_{\infty}}) $ and $\sigma$ is a constant. $\bar{A}=A/(\textbf{h}_{\infty}\sqrt{2g\textbf{h}_{\infty}})$ and $A$  is related to the radial flow velocity. $\bar{B}=B/(\textbf{h}_{\infty}\sqrt{2g\textbf{h}_{\infty}})$ and $B$ is related to the angular flow velocity. Other dimensionless parameters are $\bar{r}_{H}=r_{H}/\textbf{h}_{\infty}$, $\bar{\textbf{h}}_{H}=\textbf{h}_{H}/\textbf{h}_{\infty}$ and $\bar{g}=\tilde{g} /g$ ($\tilde{g} $ can be found in Ref. \cite{2015PhRvD..91l4018R}).

Eq. (\ref{eq. 23}) effectively rescales the parameters and variables. Provided that it is consistent with the superradiance condition given by Eq. (\ref{eq. 01}), the superradiance condition can be more intuitively expressed as
\begin{eqnarray}
0 < \tilde{\omega } < m\bar{\Omega}_{H},
\end{eqnarray}

\noindent where $\tilde{\omega }=\bar{\omega}/(\sqrt{2}/\sigma)$. All parameters and functions adhere to the conventions established in Ref. \cite{2015PhRvD..91l4018R}, with the exception that we adopt the notation $\textbf{h}$ for the free-surface function to avoid ambiguity ($\textbf{h}_{H}$ represents the water depth at the event horizon.). 

\subsection{Numerical analysis}\label{f}

Next, we will work with two distinct dimensionless parameters, which allow for a natural and immediate comparison with the Kerr black hole scattering, namely the normalized rotational flow velocity evaluated at the event horizon 
\begin{eqnarray}
\bar{v}_{\phi}=\frac{\bar{B} }{\bar{r}_{H}},
\label{eq. 24}
\end{eqnarray}

\noindent and the scaled surface gravity 
\begin{eqnarray}
\bar{\kappa}_{H}=\frac{1}{2}\frac{d}{d\bar{r}}(\bar{g}\bar{\textbf{h}}-2(\frac{\bar{A}}{\bar{r}\bar{\textbf{h}}})^2).
\label{eq. 25}
\end{eqnarray}

To facilitate an analogy with the solid material ABH model under investigation \ref{a}, we begin by defining the configuration of a rotating draining bathtub ABH, that is
\begin{eqnarray}
\textbf{h}_{\infty} \to h_{2}, \quad r_{H} \to R,
\label{eq. 26}
\end{eqnarray}

\noindent where the normalized event horizon is defined as $\bar{r}_{H} = r_{H}/\textbf{h}_{\infty}$. According to Eq. (\ref{eq. 26}), its value tends to $18.87647$. Therefore, the normalized event horizon $\bar{r}_{H}$, linear velocity $\bar{v}_{\phi}$, and surface gravity $\bar{\kappa}_{H}$ are computed by varying the normalized parameters $\bar{A}$ and $\bar{B}$ according to Eqs. (\ref{eq. 19}), (\ref{eq. 24}), (\ref{eq. 25}), and are presented in Table \ref{tab.3}. \textbf{ Parameter Settings}: Gravitational acceleration $g=9.8m/s^2$,  a tiny offset near the event horizon $\epsilon = 10^{-10}m$, and the far-field position $r_{max}=10^2m$. In addition, the factor $\sigma $ is fixed at 0.3.
\begin{table}[htbp]
	\centering
	\caption{The different values of $\bar{r}_{H}$, $\bar{v}_{\phi}$ and $\bar{\kappa}_{H}$ by changing $\bar{B}$, with fixing $\mathbf{h_{\infty}}$ and $\bar{A}$.}
	\label{tab.3}
	\renewcommand{\arraystretch}{1} 
	\setlength{\tabcolsep}{3pt} 
	\begin{tabular}{ccccccc}
		\toprule 
		No.&$\mathbf{h_{\infty}}$ (m) & $\bar{A}$ & $\bar{B}$ & $\bar{r}_{H}$&$\bar{v}_{\phi}$ &$\bar{\kappa}_{H}$\\
		\midrule 
		$A_{1}$&&&$6.44386$&$18.65709$&$0.34538$&$0.91128$\\
		$A_{2}$&&&$6.61960$&$18.80489$&$0.35202$&$0.91105$\\
		$A_{3}$&$h_{2}$	&$5.85805$&$6.73676$&$18.90434$&$0.35636$&$0.91088$\\
		$A_{4}$&&&$6.85392$&$19.00451$&$0.36065$&$0.91069$\\
		$A_{5}$&&&$7.02966$&$19.15605$&$0.36697$&$0.91038$\\
		\midrule 
		$B_{1}$&&$5.74089$&&$18.39565$&$0.35029$&$0.91977$\\
		$B_{2}$&&$5.85805$&&$18.65709$&$0.34538$&$0.91128$\\
		$B_{3}$&$h_{2}$	&$5.97521$&$6.44386$&$18.91944$&$0.34060$&$0.90301$\\
		$B_{4}$&&$6.09237$&&$19.18267$&$0.33592$&$0.89494$\\
		$B_{5}$&&$6.44386$&&$19.97749$&$0.32256$&$0.87189$\\
		\bottomrule 
	\end{tabular}
\end{table}

\begin{figure*}
	\centering 
	\includegraphics[width=0.48\textwidth]{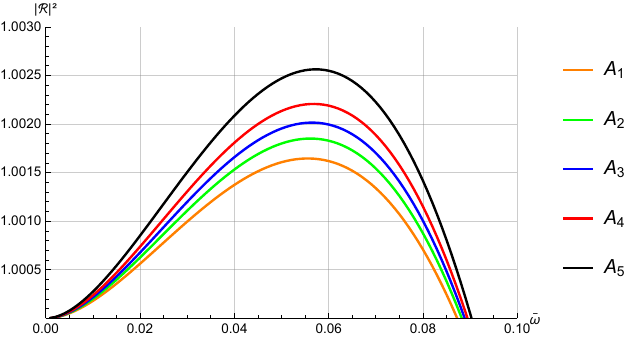}
	\includegraphics[width=0.48\textwidth]{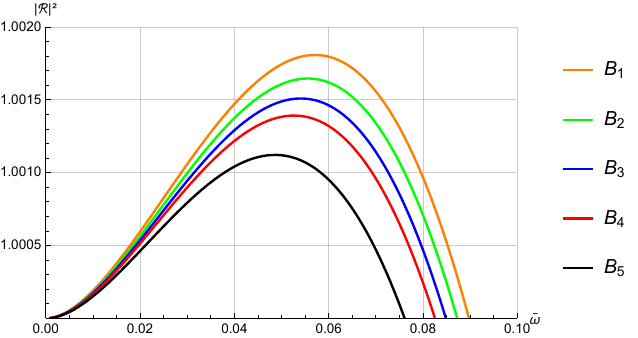}
	\includegraphics[width=0.48\textwidth]{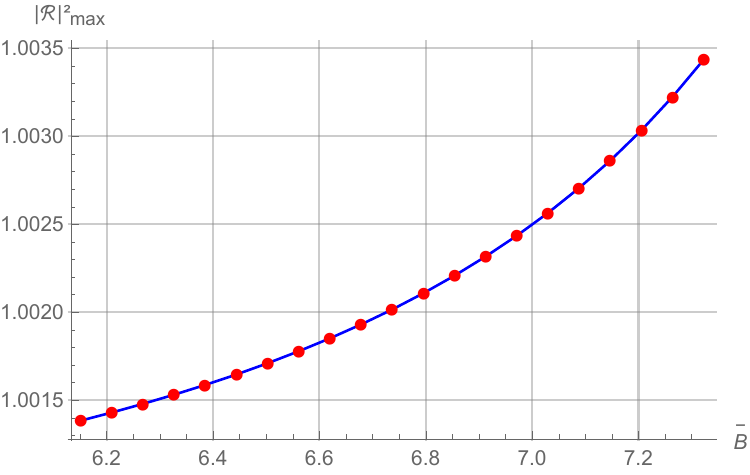}
	\includegraphics[width=0.48\textwidth]{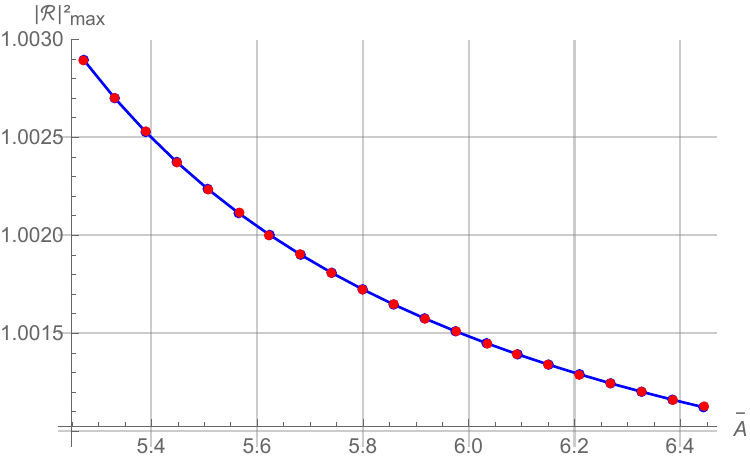}
	\caption{The top two figures illustrate the variation of the reflectivity $|\mathcal{R}|^2$ with normalized angular frequency $\bar{\omega}$, with parameters $\bar{A}$ and $\bar{B}$ held constant. Correspondingly, the bottom two figures show the maximum reflectivity $|\mathcal{R}|^2_{max}$ as a function of the normalized parameter $\bar{B}$ ($\bar{A}$).}
	\label{fig. 3}
\end{figure*}

Table \ref{tab.3} shows that when the normalized parameter $\bar{A}$ is fixed, both $\bar{r}_{H}$ and $\bar{v}_{\phi}$ exhibit a positive correlation with $\bar{B}$, whereas $\bar{\kappa}$ exhibits a negative correlation. The distinction is that, at fixed normalized parameter $\bar{B}$, only $\bar{r}_{H}$ exhibits a positive correlation with $\bar{A}$, in contrast to $\bar{v}_{\phi}$ and $\bar{\kappa}_{H}$, which exhibit a negative correlation. 
A similar pattern is also observed in the reflectivity described by Eq. (\ref{eq. 22}), which is illustrated in Fig. \ref{fig. 3}.

\subsection{Analysis and comparison}\label{g}

Fig. \ref{fig. 3} indicates that the maximum reflectivity $|\mathcal{R}|^2_{max}$ of the rotating draining bathtub ABH model exhibits a positive correlation with the normalized parameter $\bar{B}$ and a negative correlation with the normalized parameter $\bar{A}$, when the other parameter is fixed. When the water depth $\textbf{h}_{\infty}$ and event horizon of the rotating draining bathtub ABH model approach the settings of the solid material ABH model, the superradiant amplification factor reaches approximately 0.2\%—a value of the same order of magnitude as the superradiance intensity in Fig. \ref{fig. 2}. The parameter $A$ governs the radial flow velocity, analogous to the effective sound speed in the solid material ABH model. An increase in $\bar{A}$ diminishes the superradiance intensity, consistent with the suppression observed in Fig. \ref{fig. 2}.

In contrast to Eq. (\ref{eq. 01}), Eq. (\ref{eq. 23}) shows that the range of normalized angular frequencies $\bar{\omega}$ over which superradiance occurs is governed by the normalized event horizon $\bar{r}_{H}$ and the normalized parameter $\bar{B}$. Compared to the solid material ABH model and rotating draining bathtub ABH model studied in sections \ref{a} and \ref{d}, the superradiative conditions in Kerr black holes are different from the former and can be written as \cite{Andersson:1998swa}
\begin{eqnarray}
0 < \omega < \frac{ma}{2Mr_{+}}.
\end{eqnarray}

To compare all those models, we put out Tab. \ref{tab.4}. The largest possible amplification occurs when the Kerr black hole approaches extremality, i.e., $a/M\to1$. For the rotating draining bathtub ABH model, on the other hand, the maximum is attained at a sub-critical value $\bar{v}_{\phi} \approx 0.8$ \cite{2015PhRvD..91l4018R}. In the context of scalar perturbations of a nearly extremal Kerr black hole, we find that the $l = m = 2$ mode exhibits a maximum amplification coefficient of 0.2\% \cite{Andersson:1998swa}. It has also been shown that the maximum possible amplification is 0.3\% for scalar waves \cite{1972PRE}.

\begin{table*}[htbp]
	\centering
	\caption{Parameters and analogies across the solid material ABH structure model, Rotational draining bathtub ABH, and Kerr black hole for superradiance.}
	\label{tab.4}
\begin{tabular}{lccccc}
	\toprule
	& \multicolumn{4}{c}{Theoretical Models} \\
	\cmidrule{2-5}
	&\multirow{2}{*}{ \makecell{\textbf{Model I} \\ ABH structure model\\(Sec. \ref{a}\ref{b})}} 
	&\multicolumn{2}{c}{\makecell{\textbf{Model II} \\ Rotational draining \\ bathtub ABH}} 
	&\multirow{2}{*}{ \makecell{\textbf{Model III} \\ Kerr black hole\\\cite{Andersson:1998swa}}}& \\
	\cmidrule{3-4}
	Description 
	&&\makecell{\textbf{II.1}\\ Scaling\\\cite{2015PhRvD..91l4018R}}
	&\makecell{\textbf{II.2}\\ Re-scaling\\\ref{d}}
	&& Analogy\\
	\midrule
	Radial coordinate & $r$ &$r$& \makecell{$\bar{r}$\\$=r/\textbf{h}_{\infty}$ }&$r$ 
	& \makecell{\textbf{I} \& \textbf{II.1}\\ SI units\\ 
		\textbf{II.2}\\ no units\\ 
		\textbf{III}\\ \makecell{Geometrized units}\\ $(G=c=1)$}\\
		\midrule
		Event horizon & $R$ &$r_{H}$& $\bar{r}_{H}$ & $r_{+}$ &\makecell{\textbf{I} \& \textbf{II.1}\\ Consistent scale \\(Eq. (\ref{eq. 26})) }\\
		Surface gravity&/&$\kappa_{H}$&$\bar{\kappa}_{H}$&$\kappa _{+}$&\makecell{\textbf{II} \& \textbf{III}\\ Describe the ‘strength’ \\ of the horizon }\\
		\midrule
		Angular frequency & $\omega $ &$\omega $& \makecell{$\bar{\omega }$\\$=\omega /\omega_{disp}$ }& $\omega $ &\makecell{\textbf{II.2}\\ $ \omega_{disp}=\sigma \sqrt{g/\textbf{h}_{\infty}}$}\\
		Angular velocity & $\Omega $ &\makecell{$\Omega _{H}$\\$=B/r_{H}^2$}&\makecell{ $\bar{\Omega }_{H}$\\$=\bar{B}/\bar{r}^{2}_{H}$}&\makecell{$\Omega_{+}$\\$=a/(2Mr_{+})$}&\makecell{\textbf{II} \& \textbf{III}\\ Determined by the \\ rotation parameter}\\
		Condition &$\omega<m\Omega$&$\omega<m\Omega_{H}$&$\tilde{\omega } < m\bar{\Omega}_{H}$&$\omega<m\Omega_{+}$&\makecell{\textbf{I} \& \textbf{II} \& \textbf{III}\\ Consistent superradiance \\conditions}\\
		\midrule
		Wave velocity& $c_{s}$ &\makecell{$c_{\infty}$\\$=\sqrt{g\textbf{h}_{\infty}}$}& $1$ &$1$&\makecell{\textbf{I} \& \textbf{II} \& \textbf{III}\\ Constant wave speed\\ at infinity}\\
		Radial velocity &$v(r)$&\makecell{$v_{r}$\\$=A/r_{H}$}&\makecell{$\bar{v}_{r}$\\$=\bar{A}/\bar{r}_{H}$}&/&\makecell{\textbf{I} \& \textbf{II}\\Can reduce the intensity \\of superradiance (Fig. \ref{fig. 2} \ref{fig. 3})}\\
		Axial velocity&$\Omega *R$&\makecell{$v_{\phi}$\\$=B/r_{H}$}&\makecell{$\bar{v}_{\phi }$\\$=\bar{B}/\bar{r}_{H}$}&$a/M$&\makecell{\textbf{I} \& \textbf{II} \& \textbf{III}\\ Determine superradiance \\range}\\
		\midrule
		Potential function &/&\makecell{Eq. (35) \\in Ref.\cite{2015PhRvD..91l4018R}}&$\bar{V}(\bar{r})$&\makecell{Eq. (3)\\in Ref.\cite{Andersson:1998swa}}&\makecell{\textbf{II} \& \textbf{III}\\ Similar asymptotic behavior\\ (Infinity: $V\propto \omega^2$, Event \\ horizon: $V\propto (\omega-m\Omega)^2$)}\\
	     Control factors&$(\Omega, v, Z)$&$(\Omega_{H}, v_{r})$&($\bar{\Omega}_{H}, \bar{v}_{r}$)&$\Omega_{+}$&\makecell{\textbf{I} \& \textbf{II} \& \textbf{III}\\Angular velocity controls\\superradiance range(intensity)\\ \textbf{I}\& \textbf{II}\\Radial velocity weaken\\superradiance  \\ \textbf{I}\\ Impedance controls intensity}\\
		\cmidrule{2-5}
		\makecell{Superradiance \\ (Current value setting)}&\multicolumn{4}{c}{ $\sim 0.2\%$}&\makecell{\textbf{I} \& \textbf{II} \& \textbf{III}\\ Can reach the same order of\\ superradiance magnitude}\\
		\bottomrule
	\end{tabular} 
\end{table*}

For angular velocities near the critical value, the rotational draining bathtub ABH model can achieve a maximum amplification of 35\% \cite{2015PhRvD..91l4018R} when we fix $\bar{v}_{\phi }=0.8$ and $\bar{\kappa }_{H}=1.2$ and vary the depth and radius. This value is higher than the peak magnification of 21.2\% predicted by Berti's ideal-fluid model with different parameters for the rotational parameter $B = 1$ ($m = 1$) 
\cite{Berti:2004ju}. While the rotational parameter $B$ in the “draining bathtub” model has no theoretical upper bound, experimental realizations require $B\ll B_{max}$ due to fluid dynamic constraints.

\section{Conclusion}

In conclusion, our analysis has provided a first understanding of the superradiant mode in the solid material ABHs.
By examining the radial equations of ABH both inside and outside the boundary, we derive the superradiant mode formulas. Our results reveal that the internal structure of ABH significantly weakens superradiant amplification compared to a regular cylinder, due to its sound-absorbing properties. In addition, the relation of sound-wave frequency and angular velocity is investigated in simulations, and the generation of the superradiant mode in acoustic black holes is demonstrated. 
This study also demonstrates the importance of the fiber material's properties in generating the superradiant mode, and further research is needed to fully understand the properties of different materials. 
Our study has important implications for the superradiant experiments. We can observe the superradiant phenomenon by selecting appropriate acoustic materials and optimizing experimental conditions. 
Moreover,  the solid material ABH model and rotating draining bathtub ABH model show superradiance of the same magnitude at the same physical scales, aligning with extremal Kerr black hole predictions. Moreover, the solid material ABH has the most degrees of freedom. This result offers a key theoretical framework for clarifying the fundamental mechanism of superradiance in GR black holes. In summary, the combination of semi-analytical and simulation methods has allowed us to have a deeper understanding of the superradiant phenomenon in ABHs and its comparison with a regular cylinder. 
Our results have important implications for the experimental detection of the superradiant effect and provide valuable guidance for future research in this area.


\begin{thebibliography}{81}%
\makeatletter
\providecommand \@ifxundefined [1]{%
 \@ifx{#1\undefined}
}%
\providecommand \@ifnum [1]{%
 \ifnum #1\expandafter \@firstoftwo
 \else \expandafter \@secondoftwo
 \fi
}%
\providecommand \@ifx [1]{%
 \ifx #1\expandafter \@firstoftwo
 \else \expandafter \@secondoftwo
 \fi
}%
\providecommand \natexlab [1]{#1}%
\providecommand \enquote  [1]{``#1''}%
\providecommand \bibnamefont  [1]{#1}%
\providecommand \bibfnamefont [1]{#1}%
\providecommand \citenamefont [1]{#1}%
\providecommand \href@noop [0]{\@secondoftwo}%
\providecommand \href [0]{\begingroup \@sanitize@url \@href}%
\providecommand \@href[1]{\@@startlink{#1}\@@href}%
\providecommand \@@href[1]{\endgroup#1\@@endlink}%
\providecommand \@sanitize@url [0]{\catcode `\\12\catcode `\$12\catcode
  `\&12\catcode `\#12\catcode `\^12\catcode `\_12\catcode `\%12\relax}%
\providecommand \@@startlink[1]{}%
\providecommand \@@endlink[0]{}%
\providecommand \url  [0]{\begingroup\@sanitize@url \@url }%
\providecommand \@url [1]{\endgroup\@href {#1}{\urlprefix }}%
\providecommand \urlprefix  [0]{URL }%
\providecommand \Eprint [0]{\href }%
\providecommand \doibase [0]{https://doi.org/}%
\providecommand \selectlanguage [0]{\@gobble}%
\providecommand \bibinfo  [0]{\@secondoftwo}%
\providecommand \bibfield  [0]{\@secondoftwo}%
\providecommand \translation [1]{[#1]}%
\providecommand \BibitemOpen [0]{}%
\providecommand \bibitemStop [0]{}%
\providecommand \bibitemNoStop [0]{.\EOS\space}%
\providecommand \EOS [0]{\spacefactor3000\relax}%
\providecommand \BibitemShut  [1]{\csname bibitem#1\endcsname}%
\let\auto@bib@innerbib\@empty
\bibitem [{\citenamefont {Dicke}(1954)}]{PhysRev.93.99}%
  \BibitemOpen
  \bibfield  {author} {\bibinfo {author} {\bibfnamefont {R.~H.}\ \bibnamefont
  {Dicke}},\ }\bibfield  {title} {\bibinfo {title} {Coherence in spontaneous
  radiation processes},\ }\href {https://doi.org/10.1103/PhysRev.93.99}
  {\bibfield  {journal} {\bibinfo  {journal} {Phys. Rev.}\ }\textbf {\bibinfo
  {volume} {93}},\ \bibinfo {pages} {99} (\bibinfo {year} {1954})}\BibitemShut
  {NoStop}%
\bibitem [{\citenamefont {{Brito}}\ \emph
  {et~al.}(2015{\natexlab{a}})\citenamefont {{Brito}}, \citenamefont
  {{Cardoso}},\ and\ \citenamefont {{Pani}}}]{2015LNP...906.....B}%
  \BibitemOpen
  \bibfield  {author} {\bibinfo {author} {\bibfnamefont {R.}~\bibnamefont
  {{Brito}}}, \bibinfo {author} {\bibfnamefont {V.}~\bibnamefont {{Cardoso}}},\
  and\ \bibinfo {author} {\bibfnamefont {P.}~\bibnamefont {{Pani}}},\ }\href
  {https://doi.org/10.1007/978-3-030-46622-0} {\emph {\bibinfo {title}
  {{Superradiance}}}},\ Vol.\ \bibinfo {volume} {906}\ (\bibinfo {year}
  {2015})\BibitemShut {NoStop}%
\bibitem [{\citenamefont {{Penrose}}(1969)}]{1969NCimR...1..252P}%
  \BibitemOpen
  \bibfield  {author} {\bibinfo {author} {\bibfnamefont {R.}~\bibnamefont
  {{Penrose}}},\ }\bibfield  {title} {\bibinfo {title} {{Gravitational
  Collapse: the Role of General Relativity}},\ }\href@noop {} {\bibfield
  {journal} {\bibinfo  {journal} {Nuovo Cimento Rivista Serie}\ }\textbf
  {\bibinfo {volume} {1}},\ \bibinfo {pages} {252} (\bibinfo {year}
  {1969})}\BibitemShut {NoStop}%
\bibitem [{\citenamefont {Penrose}(2002)}]{2002PER}%
  \BibitemOpen
  \bibfield  {author} {\bibinfo {author} {\bibfnamefont {R.}~\bibnamefont
  {Penrose}},\ }\href {https://doi.org/10.1023/A:1016578408204} {\bibfield
  {journal} {\bibinfo  {journal} {Gen. Relat. Gravit.}\ }\textbf {\bibinfo
  {volume} {34}},\ \bibinfo {pages} {1141} (\bibinfo {year}
  {2002})}\BibitemShut {NoStop}%
\bibitem [{\citenamefont {{Zel'Dovich}}(1972)}]{1972JETP...35.1085Z}%
  \BibitemOpen
  \bibfield  {author} {\bibinfo {author} {\bibfnamefont {Y.~B.}\ \bibnamefont
  {{Zel'Dovich}}},\ }\bibfield  {title} {\bibinfo {title} {{Amplification of
  Cylindrical Electromagnetic Waves Reflected from a Rotating Body}},\
  }\href@noop {} {\bibfield  {journal} {\bibinfo  {journal} {JETP}\ }\textbf
  {\bibinfo {volume} {35}},\ \bibinfo {pages} {1085} (\bibinfo {year}
  {1972})}\BibitemShut {NoStop}%
\bibitem [{\citenamefont {{Zel'Dovich}}(1971)}]{1971JETPL..14..180Z}%
  \BibitemOpen
  \bibfield  {author} {\bibinfo {author} {\bibfnamefont {Y.~B.}\ \bibnamefont
  {{Zel'Dovich}}},\ }\bibfield  {title} {\bibinfo {title} {{Generation of Waves
  by a Rotating Body}},\ }\href@noop {} {\bibfield  {journal} {\bibinfo
  {journal} {JETP Letters}\ }\textbf {\bibinfo {volume} {14}},\ \bibinfo
  {pages} {180} (\bibinfo {year} {1971})}\BibitemShut {NoStop}%
\bibitem [{\citenamefont {{Hod}}(2016)}]{2016PhLB..758..181H}%
  \BibitemOpen
  \bibfield  {author} {\bibinfo {author} {\bibfnamefont {S.}~\bibnamefont
  {{Hod}}},\ }\bibfield  {title} {\bibinfo {title} {{The superradiant
  instability regime of the spinning Kerr black hole}},\ }\href
  {https://doi.org/10.1016/j.physletb.2016.05.012} {\bibfield  {journal}
  {\bibinfo  {journal} {Phys. Lett. B}\ }\textbf {\bibinfo {volume} {758}},\
  \bibinfo {pages} {181} (\bibinfo {year} {2016})}\BibitemShut {NoStop}%
\bibitem [{\citenamefont {{Arvanitaki}}\ \emph {et~al.}(2010)\citenamefont
  {{Arvanitaki}}, \citenamefont {{Dimopoulos}}, \citenamefont {{Dubovsky}},
  \citenamefont {{Kaloper}},\ and\ \citenamefont
  {{March-Russell}}}]{2010PhRvD..81l3530A}%
  \BibitemOpen
  \bibfield  {author} {\bibinfo {author} {\bibfnamefont {A.}~\bibnamefont
  {{Arvanitaki}}}, \bibinfo {author} {\bibfnamefont {S.}~\bibnamefont
  {{Dimopoulos}}}, \bibinfo {author} {\bibfnamefont {S.}~\bibnamefont
  {{Dubovsky}}}, \bibinfo {author} {\bibfnamefont {N.}~\bibnamefont
  {{Kaloper}}},\ and\ \bibinfo {author} {\bibfnamefont {J.}~\bibnamefont
  {{March-Russell}}},\ }\bibfield  {title} {\bibinfo {title} {{String
  axiverse}},\ }\href {https://doi.org/10.1103/PhysRevD.81.123530} {\bibfield
  {journal} {\bibinfo  {journal} {\prd}\ }\textbf {\bibinfo {volume} {81}},\
  \bibinfo {eid} {123530} (\bibinfo {year} {2010})}\BibitemShut {NoStop}%
\bibitem [{\citenamefont {{Arvanitaki}}\ and\ \citenamefont
  {{Dubovsky}}(2011)}]{2011PhRvD..83d4026A}%
  \BibitemOpen
  \bibfield  {author} {\bibinfo {author} {\bibfnamefont {A.}~\bibnamefont
  {{Arvanitaki}}}\ and\ \bibinfo {author} {\bibfnamefont {S.}~\bibnamefont
  {{Dubovsky}}},\ }\bibfield  {title} {\bibinfo {title} {{Exploring the string
  axiverse with precision black hole physics}},\ }\href
  {https://doi.org/10.1103/PhysRevD.83.044026} {\bibfield  {journal} {\bibinfo
  {journal} {\prd}\ }\textbf {\bibinfo {volume} {83}},\ \bibinfo {eid} {044026}
  (\bibinfo {year} {2011})}\BibitemShut {NoStop}%
\bibitem [{\citenamefont {{Andersson}}\ and\ \citenamefont
  {{Glampedakis}}(2000)}]{2000PhRvL..84.4537A}%
  \BibitemOpen
  \bibfield  {author} {\bibinfo {author} {\bibfnamefont {N.}~\bibnamefont
  {{Andersson}}}\ and\ \bibinfo {author} {\bibfnamefont {K.}~\bibnamefont
  {{Glampedakis}}},\ }\bibfield  {title} {\bibinfo {title} {{Superradiance
  Resonance Cavity Outside Rapidly Rotating Black Holes}},\ }\href
  {https://doi.org/10.1103/PhysRevLett.84.4537} {\bibfield  {journal} {\bibinfo
   {journal} {\prl}\ }\textbf {\bibinfo {volume} {84}},\ \bibinfo {pages}
  {4537} (\bibinfo {year} {2000})}\BibitemShut {NoStop}%
\bibitem [{\citenamefont {{Cardoso}}\ and\ \citenamefont
  {{Dias}}(2004)}]{2004PhRvD..70h4011C}%
  \BibitemOpen
  \bibfield  {author} {\bibinfo {author} {\bibfnamefont {V.}~\bibnamefont
  {{Cardoso}}}\ and\ \bibinfo {author} {\bibfnamefont {{\'O}.~J.}\ \bibnamefont
  {{Dias}}},\ }\bibfield  {title} {\bibinfo {title} {{Small Kerr anti-de Sitter
  black holes are unstable}},\ }\href
  {https://doi.org/10.1103/PhysRevD.70.084011} {\bibfield  {journal} {\bibinfo
  {journal} {\prd}\ }\textbf {\bibinfo {volume} {70}},\ \bibinfo {eid} {084011}
  (\bibinfo {year} {2004})}\BibitemShut {NoStop}%
\bibitem [{\citenamefont {{Witek}}\ \emph {et~al.}(2013)\citenamefont
  {{Witek}}, \citenamefont {{Cardoso}}, \citenamefont {{Ishibashi}},\ and\
  \citenamefont {{Sperhake}}}]{2013PhRvD..87d3513W}%
  \BibitemOpen
  \bibfield  {author} {\bibinfo {author} {\bibfnamefont {H.}~\bibnamefont
  {{Witek}}}, \bibinfo {author} {\bibfnamefont {V.}~\bibnamefont {{Cardoso}}},
  \bibinfo {author} {\bibfnamefont {A.}~\bibnamefont {{Ishibashi}}},\ and\
  \bibinfo {author} {\bibfnamefont {U.}~\bibnamefont {{Sperhake}}},\ }\bibfield
   {title} {\bibinfo {title} {{Superradiant instabilities in astrophysical
  systems}},\ }\href {https://doi.org/10.1103/PhysRevD.87.043513} {\bibfield
  {journal} {\bibinfo  {journal} {\prd}\ }\textbf {\bibinfo {volume} {87}},\
  \bibinfo {eid} {043513} (\bibinfo {year} {2013})}\BibitemShut {NoStop}%
\bibitem [{\citenamefont {{Mehta}}\ \emph {et~al.}(2021)\citenamefont
  {{Mehta}}, \citenamefont {{Demirtas}}, \citenamefont {{Long}}, \citenamefont
  {{Marsh}}, \citenamefont {{McAllister}},\ and\ \citenamefont
  {{Stott}}}]{2021JCAP...07..033M}%
  \BibitemOpen
  \bibfield  {author} {\bibinfo {author} {\bibfnamefont {V.~M.}\ \bibnamefont
  {{Mehta}}}, \bibinfo {author} {\bibfnamefont {M.}~\bibnamefont {{Demirtas}}},
  \bibinfo {author} {\bibfnamefont {C.}~\bibnamefont {{Long}}}, \bibinfo
  {author} {\bibfnamefont {D.~J.~E.}\ \bibnamefont {{Marsh}}}, \bibinfo
  {author} {\bibfnamefont {L.}~\bibnamefont {{McAllister}}},\ and\ \bibinfo
  {author} {\bibfnamefont {M.~J.}\ \bibnamefont {{Stott}}},\ }\bibfield
  {title} {\bibinfo {title} {{Superradiance in string theory}},\ }\href
  {https://doi.org/10.1088/1475-7516/2021/07/033} {\bibfield  {journal}
  {\bibinfo  {journal} {JCAP}\ }\textbf {\bibinfo {volume} {2021}}\bibinfo
  {number} { (7)},\ \bibinfo {eid} {033}}\BibitemShut {NoStop}%
\bibitem [{\citenamefont {{Baryakhtar}}\ \emph {et~al.}(2017)\citenamefont
  {{Baryakhtar}}, \citenamefont {{Lasenby}},\ and\ \citenamefont
  {{Teo}}}]{2017PhRvD..96c5019B}%
  \BibitemOpen
\bibfield  {number} {  }\bibfield  {author} {\bibinfo {author} {\bibfnamefont
  {M.}~\bibnamefont {{Baryakhtar}}}, \bibinfo {author} {\bibfnamefont
  {R.}~\bibnamefont {{Lasenby}}},\ and\ \bibinfo {author} {\bibfnamefont
  {M.}~\bibnamefont {{Teo}}},\ }\bibfield  {title} {\bibinfo {title} {{Black
  hole superradiance signatures of ultralight vectors}},\ }\href
  {https://doi.org/10.1103/PhysRevD.96.035019} {\bibfield  {journal} {\bibinfo
  {journal} {\prd}\ }\textbf {\bibinfo {volume} {96}},\ \bibinfo {eid} {035019}
  (\bibinfo {year} {2017})}\BibitemShut {NoStop}%
\bibitem [{\citenamefont {Degollado}\ \emph {et~al.}(2013)\citenamefont
  {Degollado}, \citenamefont {Herdeiro},\ and\ \citenamefont
  {R\'unarsson}}]{PhysRevD.88.063003}%
  \BibitemOpen
  \bibfield  {author} {\bibinfo {author} {\bibfnamefont {J.~C.}\ \bibnamefont
  {Degollado}}, \bibinfo {author} {\bibfnamefont {C.~A.~R.}\ \bibnamefont
  {Herdeiro}},\ and\ \bibinfo {author} {\bibfnamefont {H.~F.}\ \bibnamefont
  {R\'unarsson}},\ }\bibfield  {title} {\bibinfo {title} {Rapid growth of
  superradiant instabilities for charged black holes in a cavity},\ }\href
  {https://doi.org/10.1103/PhysRevD.88.063003} {\bibfield  {journal} {\bibinfo
  {journal} {Phys. Rev. D}\ }\textbf {\bibinfo {volume} {88}},\ \bibinfo
  {pages} {063003} (\bibinfo {year} {2013})}\BibitemShut {NoStop}%
\bibitem [{\citenamefont {{Degollado}}\ \emph {et~al.}(2018)\citenamefont
  {{Degollado}}, \citenamefont {{Herdeiro}},\ and\ \citenamefont
  {{Radu}}}]{2018PhLB..781..651D}%
  \BibitemOpen
  \bibfield  {author} {\bibinfo {author} {\bibfnamefont {J.~C.}\ \bibnamefont
  {{Degollado}}}, \bibinfo {author} {\bibfnamefont {C.~A.~R.}\ \bibnamefont
  {{Herdeiro}}},\ and\ \bibinfo {author} {\bibfnamefont {E.}~\bibnamefont
  {{Radu}}},\ }\bibfield  {title} {\bibinfo {title} {{Effective stability
  against superradiance of Kerr black holes with synchronised hair}},\ }\href
  {https://doi.org/10.1016/j.physletb.2018.04.052} {\bibfield  {journal}
  {\bibinfo  {journal} {Phys. Lett. B}\ }\textbf {\bibinfo {volume} {781}},\
  \bibinfo {pages} {651} (\bibinfo {year} {2018})}\BibitemShut {NoStop}%
\bibitem [{\citenamefont {{Brito}}\ \emph
  {et~al.}(2015{\natexlab{b}})\citenamefont {{Brito}}, \citenamefont
  {{Cardoso}},\ and\ \citenamefont {{Pani}}}]{2015CQGra..32m4001B}%
  \BibitemOpen
  \bibfield  {author} {\bibinfo {author} {\bibfnamefont {R.}~\bibnamefont
  {{Brito}}}, \bibinfo {author} {\bibfnamefont {V.}~\bibnamefont {{Cardoso}}},\
  and\ \bibinfo {author} {\bibfnamefont {P.}~\bibnamefont {{Pani}}},\
  }\bibfield  {title} {\bibinfo {title} {{Black holes as particle detectors:
  evolution of superradiant instabilities}},\ }\href
  {https://doi.org/10.1088/0264-9381/32/13/134001} {\bibfield  {journal}
  {\bibinfo  {journal} {Class. Quant. Grav.}\ }\textbf {\bibinfo {volume}
  {32}},\ \bibinfo {eid} {134001} (\bibinfo {year}
  {2015}{\natexlab{b}})}\BibitemShut {NoStop}%
\bibitem [{\citenamefont {{Ghosh}}\ \emph {et~al.}(2019)\citenamefont
  {{Ghosh}}, \citenamefont {{Berti}}, \citenamefont {{Brito}},\ and\
  \citenamefont {{Richartz}}}]{2019PhRvD..99j4030G}%
  \BibitemOpen
  \bibfield  {author} {\bibinfo {author} {\bibfnamefont {S.}~\bibnamefont
  {{Ghosh}}}, \bibinfo {author} {\bibfnamefont {E.}~\bibnamefont {{Berti}}},
  \bibinfo {author} {\bibfnamefont {R.}~\bibnamefont {{Brito}}},\ and\ \bibinfo
  {author} {\bibfnamefont {M.}~\bibnamefont {{Richartz}}},\ }\bibfield  {title}
  {\bibinfo {title} {{Follow-up signals from superradiant instabilities of
  black hole merger remnants}},\ }\href
  {https://doi.org/10.1103/PhysRevD.99.104030} {\bibfield  {journal} {\bibinfo
  {journal} {\prd}\ }\textbf {\bibinfo {volume} {99}},\ \bibinfo {eid} {104030}
  (\bibinfo {year} {2019})}\BibitemShut {NoStop}%
\bibitem [{\citenamefont {{Casals}}\ \emph {et~al.}(2008)\citenamefont
  {{Casals}}, \citenamefont {{Dolan}}, \citenamefont {{Kanti}},\ and\
  \citenamefont {{Winstanley}}}]{2008JHEP...06..071C}%
  \BibitemOpen
  \bibfield  {author} {\bibinfo {author} {\bibfnamefont {M.}~\bibnamefont
  {{Casals}}}, \bibinfo {author} {\bibfnamefont {S.}~\bibnamefont {{Dolan}}},
  \bibinfo {author} {\bibfnamefont {P.}~\bibnamefont {{Kanti}}},\ and\ \bibinfo
  {author} {\bibfnamefont {E.}~\bibnamefont {{Winstanley}}},\ }\bibfield
  {title} {\bibinfo {title} {{Bulk emission of scalars by a rotating black
  hole}},\ }\href {https://doi.org/10.1088/1126-6708/2008/06/071} {\bibfield
  {journal} {\bibinfo  {journal} {JHEP}\ }\textbf {\bibinfo {volume}
  {2008}}\bibinfo  {number} { (6)},\ \bibinfo {eid} {071}}\BibitemShut
  {NoStop}%
\bibitem [{\citenamefont {{Rosa}}\ and\ \citenamefont
  {{Kephart}}(2018)}]{2018PhRvL.120w1102R}%
  \BibitemOpen
\bibfield  {number} {  }\bibfield  {author} {\bibinfo {author} {\bibfnamefont
  {J.~G.}\ \bibnamefont {{Rosa}}}\ and\ \bibinfo {author} {\bibfnamefont
  {T.~W.}\ \bibnamefont {{Kephart}}},\ }\bibfield  {title} {\bibinfo {title}
  {{Stimulated Axion Decay in Superradiant Clouds around Primordial Black
  Holes}},\ }\href {https://doi.org/10.1103/PhysRevLett.120.231102} {\bibfield
  {journal} {\bibinfo  {journal} {\prl}\ }\textbf {\bibinfo {volume} {120}},\
  \bibinfo {eid} {231102} (\bibinfo {year} {2018})}\BibitemShut {NoStop}%
\bibitem [{\citenamefont {{East}}\ and\ \citenamefont
  {{Pretorius}}(2017)}]{2017PhRvL.119d1101E}%
  \BibitemOpen
  \bibfield  {author} {\bibinfo {author} {\bibfnamefont {W.~E.}\ \bibnamefont
  {{East}}}\ and\ \bibinfo {author} {\bibfnamefont {F.}~\bibnamefont
  {{Pretorius}}},\ }\bibfield  {title} {\bibinfo {title} {{Superradiant
  Instability and Backreaction of Massive Vector Fields around Kerr Black
  Holes}},\ }\href {https://doi.org/10.1103/PhysRevLett.119.041101} {\bibfield
  {journal} {\bibinfo  {journal} {\prl}\ }\textbf {\bibinfo {volume} {119}},\
  \bibinfo {eid} {041101} (\bibinfo {year} {2017})}\BibitemShut {NoStop}%
\bibitem [{\citenamefont {{Konoplya}}\ and\ \citenamefont
  {{Zhidenko}}(2016)}]{2016JCAP...12..043K}%
  \BibitemOpen
  \bibfield  {author} {\bibinfo {author} {\bibfnamefont {R.~A.}\ \bibnamefont
  {{Konoplya}}}\ and\ \bibinfo {author} {\bibfnamefont {A.}~\bibnamefont
  {{Zhidenko}}},\ }\bibfield  {title} {\bibinfo {title} {{Wormholes versus
  black holes: quasinormal ringing at early and late times}},\ }\href
  {https://doi.org/10.1088/1475-7516/2016/12/043} {\bibfield  {journal}
  {\bibinfo  {journal} {JCAP}\ }\textbf {\bibinfo {volume} {2016}}\bibinfo
  {number} { (12)},\ \bibinfo {eid} {043}}\BibitemShut {NoStop}%
\bibitem [{\citenamefont {{Richartz}}\ \emph {et~al.}(2015)\citenamefont
  {{Richartz}}, \citenamefont {{Prain}}, \citenamefont {{Liberati}},\ and\
  \citenamefont {{Weinfurtner}}}]{2015PhRvD..91l4018R}%
  \BibitemOpen
\bibfield  {number} {  }\bibfield  {author} {\bibinfo {author} {\bibfnamefont
  {M.}~\bibnamefont {{Richartz}}}, \bibinfo {author} {\bibfnamefont
  {A.}~\bibnamefont {{Prain}}}, \bibinfo {author} {\bibfnamefont
  {S.}~\bibnamefont {{Liberati}}},\ and\ \bibinfo {author} {\bibfnamefont
  {S.}~\bibnamefont {{Weinfurtner}}},\ }\bibfield  {title} {\bibinfo {title}
  {{Rotating black holes in a draining bathtub: Superradiant scattering of
  gravity waves}},\ }\href {https://doi.org/10.1103/PhysRevD.91.124018}
  {\bibfield  {journal} {\bibinfo  {journal} {\prd}\ }\textbf {\bibinfo
  {volume} {91}},\ \bibinfo {eid} {124018} (\bibinfo {year}
  {2015})}\BibitemShut {NoStop}%
\bibitem [{\citenamefont {{Wang}}\ and\ \citenamefont
  {{Herdeiro}}(2016)}]{2016PhRvD..93f4066W}%
  \BibitemOpen
  \bibfield  {author} {\bibinfo {author} {\bibfnamefont {M.}~\bibnamefont
  {{Wang}}}\ and\ \bibinfo {author} {\bibfnamefont {C.}~\bibnamefont
  {{Herdeiro}}},\ }\bibfield  {title} {\bibinfo {title} {{Maxwell perturbations
  on Kerr-anti-de Sitter black holes: Quasinormal modes, superradiant
  instabilities, and vector clouds}},\ }\href
  {https://doi.org/10.1103/PhysRevD.93.064066} {\bibfield  {journal} {\bibinfo
  {journal} {\prd}\ }\textbf {\bibinfo {volume} {93}},\ \bibinfo {eid} {064066}
  (\bibinfo {year} {2016})}\BibitemShut {NoStop}%
\bibitem [{\citenamefont {{Yoshino}}\ and\ \citenamefont
  {{Kodama}}(2014)}]{2014PTEP.2014d3E02Y}%
  \BibitemOpen
  \bibfield  {author} {\bibinfo {author} {\bibfnamefont {H.}~\bibnamefont
  {{Yoshino}}}\ and\ \bibinfo {author} {\bibfnamefont {H.}~\bibnamefont
  {{Kodama}}},\ }\bibfield  {title} {\bibinfo {title} {{Gravitational radiation
  from an axion cloud around a black hole: Superradiant phase}},\ }\href
  {https://doi.org/10.1093/ptep/ptu029} {\bibfield  {journal} {\bibinfo
  {journal} {PTEP}\ }\textbf {\bibinfo {volume} {2014}},\ \bibinfo {eid}
  {043E02} (\bibinfo {year} {2014})}\BibitemShut {NoStop}%
\bibitem [{\citenamefont {{Zhang}}\ \emph {et~al.}(2020)\citenamefont
  {{Zhang}}, \citenamefont {{Zhang}}, \citenamefont {{Li}},\ and\ \citenamefont
  {{Guo}}}]{2020JHEP...08..105Z}%
  \BibitemOpen
  \bibfield  {author} {\bibinfo {author} {\bibfnamefont {C.-Y.}\ \bibnamefont
  {{Zhang}}}, \bibinfo {author} {\bibfnamefont {S.-J.}\ \bibnamefont
  {{Zhang}}}, \bibinfo {author} {\bibfnamefont {P.-C.}\ \bibnamefont {{Li}}},\
  and\ \bibinfo {author} {\bibfnamefont {M.}~\bibnamefont {{Guo}}},\ }\bibfield
   {title} {\bibinfo {title} {{Superradiance and stability of the regularized
  4D charged Einstein-Gauss-Bonnet black hole}},\ }\href
  {https://doi.org/10.1007/JHEP08(2020)105} {\bibfield  {journal} {\bibinfo
  {journal} {JHEP}\ }\textbf {\bibinfo {volume} {2020}}\bibinfo  {number} {
  (8)},\ \bibinfo {eid} {105}}\BibitemShut {NoStop}%
\bibitem [{\citenamefont {{Sun}}\ and\ \citenamefont
  {{Zhang}}(2021)}]{2021PhRvD.104j3009S}%
  \BibitemOpen
\bibfield  {number} {  }\bibfield  {author} {\bibinfo {author} {\bibfnamefont
  {S.~C.}\ \bibnamefont {{Sun}}}\ and\ \bibinfo {author} {\bibfnamefont
  {Y.~L.}\ \bibnamefont {{Zhang}}},\ }\bibfield  {title} {\bibinfo {title}
  {{Fast gravitational wave bursts from axion clumps}},\ }\href
  {https://doi.org/10.1103/PhysRevD.104.103009} {\bibfield  {journal} {\bibinfo
   {journal} {\prd}\ }\textbf {\bibinfo {volume} {104}},\ \bibinfo {eid}
  {103009} (\bibinfo {year} {2021})}\BibitemShut {NoStop}%
\bibitem [{\citenamefont {{Aggarwal}}\ \emph {et~al.}(2021)\citenamefont
  {{Aggarwal}}, \citenamefont {{Aguiar}},\ and\ \citenamefont
  {{Bauswein}}}]{2021LRR....24....4A}%
  \BibitemOpen
  \bibfield  {author} {\bibinfo {author} {\bibfnamefont {N.}~\bibnamefont
  {{Aggarwal}}}, \bibinfo {author} {\bibfnamefont {O.~D.}\ \bibnamefont
  {{Aguiar}}},\ and\ \bibinfo {author} {\bibfnamefont {A.}~\bibnamefont
  {{Bauswein}}},\ }\bibfield  {title} {\bibinfo {title} {{Challenges and
  opportunities of gravitational-wave searches at MHz to GHz frequencies}},\
  }\href {https://doi.org/10.1007/s41114-021-00032-5} {\bibfield  {journal}
  {\bibinfo  {journal} {Living Rev. Rel.}\ }\textbf {\bibinfo {volume} {24}},\
  \bibinfo {eid} {4} (\bibinfo {year} {2021})}\BibitemShut {NoStop}%
\bibitem [{\citenamefont {{Endlich}}\ and\ \citenamefont
  {{Penco}}(2017)}]{2017JHEP...05..052E}%
  \BibitemOpen
  \bibfield  {author} {\bibinfo {author} {\bibfnamefont {S.}~\bibnamefont
  {{Endlich}}}\ and\ \bibinfo {author} {\bibfnamefont {R.}~\bibnamefont
  {{Penco}}},\ }\bibfield  {title} {\bibinfo {title} {{A modern approach to
  superradiance}},\ }\href {https://doi.org/10.1007/JHEP05(2017)052} {\bibfield
   {journal} {\bibinfo  {journal} {JHEP}\ }\textbf {\bibinfo {volume}
  {2017}}\bibinfo  {number} { (5)},\ \bibinfo {eid} {52}}\BibitemShut {NoStop}%
\bibitem [{\citenamefont {{Cardoso}}\ \emph {et~al.}(2016)\citenamefont
  {{Cardoso}}, \citenamefont {{Coutant}}, \citenamefont {{Richartz}},\ and\
  \citenamefont {{Weinfurtner}}}]{2016PhRvL.117A1101C}%
  \BibitemOpen
\bibfield  {number} {  }\bibfield  {author} {\bibinfo {author} {\bibfnamefont
  {V.}~\bibnamefont {{Cardoso}}}, \bibinfo {author} {\bibfnamefont
  {A.}~\bibnamefont {{Coutant}}}, \bibinfo {author} {\bibfnamefont
  {M.}~\bibnamefont {{Richartz}}},\ and\ \bibinfo {author} {\bibfnamefont
  {S.}~\bibnamefont {{Weinfurtner}}},\ }\bibfield  {title} {\bibinfo {title}
  {{Detecting Rotational Superradiance in Fluid Laboratories}},\ }\href
  {https://doi.org/10.1103/PhysRevLett.117.271101} {\bibfield  {journal}
  {\bibinfo  {journal} {\prl}\ }\textbf {\bibinfo {volume} {117}},\ \bibinfo
  {eid} {271101} (\bibinfo {year} {2016})}\BibitemShut {NoStop}%
\bibitem [{\citenamefont {Starobinsky}(1973)}]{Starobinsky:1973aij}%
  \BibitemOpen
  \bibfield  {author} {\bibinfo {author} {\bibfnamefont {A.~A.}\ \bibnamefont
  {Starobinsky}},\ }\bibfield  {title} {\bibinfo {title} {{Amplification of
  waves reflected from a rotating ''black hole''.}},\ }\href@noop {} {\bibfield
   {journal} {\bibinfo  {journal} {Sov. Phys. JETP}\ }\textbf {\bibinfo
  {volume} {37}},\ \bibinfo {pages} {28} (\bibinfo {year} {1973})}\BibitemShut
  {NoStop}%
\bibitem [{\citenamefont {{Bekenstein}}\ and\ \citenamefont
  {{Schiffer}}(1998)}]{1998PhRvD..58f4014B}%
  \BibitemOpen
  \bibfield  {author} {\bibinfo {author} {\bibfnamefont {J.~D.}\ \bibnamefont
  {{Bekenstein}}}\ and\ \bibinfo {author} {\bibfnamefont {M.}~\bibnamefont
  {{Schiffer}}},\ }\bibfield  {title} {\bibinfo {title} {{The many faces of
  superradiance}},\ }\href {https://doi.org/10.1103/PhysRevD.58.064014}
  {\bibfield  {journal} {\bibinfo  {journal} {\prd}\ }\textbf {\bibinfo
  {volume} {58}},\ \bibinfo {eid} {064014} (\bibinfo {year}
  {1998})}\BibitemShut {NoStop}%
\bibitem [{\citenamefont {Faccio}\ and\ \citenamefont
  {Wright}(2019)}]{PhysRevLett.123.044301}%
  \BibitemOpen
  \bibfield  {author} {\bibinfo {author} {\bibfnamefont {D.}~\bibnamefont
  {Faccio}}\ and\ \bibinfo {author} {\bibfnamefont {E.~M.}\ \bibnamefont
  {Wright}},\ }\bibfield  {title} {\bibinfo {title} {Superradiant amplification
  of acoustic beams via medium rotation},\ }\href
  {https://doi.org/10.1103/PhysRevLett.123.044301} {\bibfield  {journal}
  {\bibinfo  {journal} {Phys. Rev. Lett.}\ }\textbf {\bibinfo {volume} {123}},\
  \bibinfo {pages} {044301} (\bibinfo {year} {2019})}\BibitemShut {NoStop}%
\bibitem [{\citenamefont {Torres}\ \emph {et~al.}(2017)\citenamefont {Torres},
  \citenamefont {Patrick}, \citenamefont {Coutant}, \citenamefont {Richartz},
  \citenamefont {Tedford},\ and\ \citenamefont {Weinfurtner}}]{2017TTP}%
  \BibitemOpen
  \bibfield  {author} {\bibinfo {author} {\bibfnamefont {T.}~\bibnamefont
  {Torres}}, \bibinfo {author} {\bibfnamefont {S.}~\bibnamefont {Patrick}},
  \bibinfo {author} {\bibfnamefont {A.}~\bibnamefont {Coutant}}, \bibinfo
  {author} {\bibfnamefont {M.}~\bibnamefont {Richartz}}, \bibinfo {author}
  {\bibfnamefont {E.~W.}\ \bibnamefont {Tedford}},\ and\ \bibinfo {author}
  {\bibfnamefont {S.}~\bibnamefont {Weinfurtner}},\ }\href
  {https://doi.org/10.1038/nphys4151} {\bibfield  {journal} {\bibinfo
  {journal} {Nat. Phys.}\ }\textbf {\bibinfo {volume} {13}},\ \bibinfo {pages}
  {833} (\bibinfo {year} {2017})}\BibitemShut {NoStop}%
\bibitem [{\citenamefont {Andersson}\ \emph {et~al.}(1998)\citenamefont
  {Andersson}, \citenamefont {Laguna},\ and\ \citenamefont
  {Papadopoulos}}]{Andersson:1998swa}%
  \BibitemOpen
  \bibfield  {author} {\bibinfo {author} {\bibfnamefont {N.}~\bibnamefont
  {Andersson}}, \bibinfo {author} {\bibfnamefont {P.}~\bibnamefont {Laguna}},\
  and\ \bibinfo {author} {\bibfnamefont {P.}~\bibnamefont {Papadopoulos}},\
  }\bibfield  {title} {\bibinfo {title} {{Dynamics of scalar fields in the
  background of rotating black holes. 2. A Note on superradiance}},\ }\href
  {https://doi.org/10.1103/PhysRevD.58.087503} {\bibfield  {journal} {\bibinfo
  {journal} {Phys. Rev. D}\ }\textbf {\bibinfo {volume} {58}},\ \bibinfo
  {pages} {087503} (\bibinfo {year} {1998})},\ \Eprint
  {https://arxiv.org/abs/gr-qc/9802059} {arXiv:gr-qc/9802059} \BibitemShut
  {NoStop}%
\bibitem [{\citenamefont {{Ginzburg}}\ and\ \citenamefont
  {{Frank}}(1945)}]{1945gvl}%
  \BibitemOpen
  \bibfield  {author} {\bibinfo {author} {\bibfnamefont {V.~L.}\ \bibnamefont
  {{Ginzburg}}}\ and\ \bibinfo {author} {\bibfnamefont {I.~M.}\ \bibnamefont
  {{Frank}}},\ }\href@noop {} {\bibfield  {journal} {\bibinfo  {journal} {J.
  Phys. (USSR)}\ }\textbf {\bibinfo {volume} {9}},\ \bibinfo {pages} {353}
  (\bibinfo {year} {1945})}\BibitemShut {NoStop}%
\bibitem [{\citenamefont {Cardoso}\ and\ \citenamefont
  {Yoshida}(2005)}]{Cardoso:2005vk}%
  \BibitemOpen
  \bibfield  {author} {\bibinfo {author} {\bibfnamefont {V.}~\bibnamefont
  {Cardoso}}\ and\ \bibinfo {author} {\bibfnamefont {S.}~\bibnamefont
  {Yoshida}},\ }\bibfield  {title} {\bibinfo {title} {{Superradiant
  instabilities of rotating black branes and strings}},\ }\href
  {https://doi.org/10.1088/1126-6708/2005/07/009} {\bibfield  {journal}
  {\bibinfo  {journal} {JHEP}\ }\textbf {\bibinfo {volume} {07}},\ \bibinfo
  {pages} {009}},\ \Eprint {https://arxiv.org/abs/hep-th/0502206}
  {arXiv:hep-th/0502206} \BibitemShut {NoStop}%
\bibitem [{\citenamefont {{Saffin}}\ \emph {et~al.}(2023)\citenamefont
  {{Saffin}}, \citenamefont {{Xie}},\ and\ \citenamefont
  {{Zhou}}}]{2023PhRvL.131k1601S}%
  \BibitemOpen
  \bibfield  {author} {\bibinfo {author} {\bibfnamefont {P.~M.}\ \bibnamefont
  {{Saffin}}}, \bibinfo {author} {\bibfnamefont {Q.~X.}\ \bibnamefont
  {{Xie}}},\ and\ \bibinfo {author} {\bibfnamefont {S.~Y.}\ \bibnamefont
  {{Zhou}}},\ }\bibfield  {title} {\bibinfo {title} {{Q -Ball Superradiance}},\
  }\href {https://doi.org/10.1103/PhysRevLett.131.111601} {\bibfield  {journal}
  {\bibinfo  {journal} {\prl}\ }\textbf {\bibinfo {volume} {131}},\ \bibinfo
  {eid} {111601} (\bibinfo {year} {2023})}\BibitemShut {NoStop}%
\bibitem [{\citenamefont {{Yang}}\ \emph {et~al.}(2021)\citenamefont {{Yang}},
  \citenamefont {{Oh}}, \citenamefont {{Han}}, \citenamefont {{Son}},
  \citenamefont {{Kim}}, \citenamefont {{Kim}}, \citenamefont {{Lee}},\ and\
  \citenamefont {{An}}}]{2021NaPho..15..272Y}%
  \BibitemOpen
  \bibfield  {author} {\bibinfo {author} {\bibfnamefont {D.}~\bibnamefont
  {{Yang}}}, \bibinfo {author} {\bibfnamefont {S.}~\bibnamefont {{Oh}}},
  \bibinfo {author} {\bibfnamefont {J.}~\bibnamefont {{Han}}}, \bibinfo
  {author} {\bibfnamefont {G.}~\bibnamefont {{Son}}}, \bibinfo {author}
  {\bibfnamefont {J.}~\bibnamefont {{Kim}}}, \bibinfo {author} {\bibfnamefont
  {J.}~\bibnamefont {{Kim}}}, \bibinfo {author} {\bibfnamefont
  {M.}~\bibnamefont {{Lee}}},\ and\ \bibinfo {author} {\bibfnamefont
  {K.}~\bibnamefont {{An}}},\ }\bibfield  {title} {\bibinfo {title}
  {{Realization of superabsorption by time reversal of superradiance}},\ }\href
  {https://doi.org/10.1038/s41566-021-00770-6} {\bibfield  {journal} {\bibinfo
  {journal} {Nat. Photon.}\ }\textbf {\bibinfo {volume} {15}},\ \bibinfo
  {pages} {272} (\bibinfo {year} {2021})}\BibitemShut {NoStop}%
\bibitem [{\citenamefont {{Kim}}\ \emph {et~al.}(2022)\citenamefont {{Kim}},
  \citenamefont {{Oh}}, \citenamefont {{Yang}}, \citenamefont {{Kim}},
  \citenamefont {{Lee}},\ and\ \citenamefont {{An}}}]{2022NaPho..16..707K}%
  \BibitemOpen
  \bibfield  {author} {\bibinfo {author} {\bibfnamefont {J.}~\bibnamefont
  {{Kim}}}, \bibinfo {author} {\bibfnamefont {S.}~\bibnamefont {{Oh}}},
  \bibinfo {author} {\bibfnamefont {D.}~\bibnamefont {{Yang}}}, \bibinfo
  {author} {\bibfnamefont {J.}~\bibnamefont {{Kim}}}, \bibinfo {author}
  {\bibfnamefont {M.}~\bibnamefont {{Lee}}},\ and\ \bibinfo {author}
  {\bibfnamefont {K.}~\bibnamefont {{An}}},\ }\bibfield  {title} {\bibinfo
  {title} {{A photonic quantum engine driven by superradiance}},\ }\href
  {https://doi.org/10.1038/s41566-022-01039-2} {\bibfield  {journal} {\bibinfo
  {journal} {Nat. Photon.}\ }\textbf {\bibinfo {volume} {16}},\ \bibinfo
  {pages} {707} (\bibinfo {year} {2022})}\BibitemShut {NoStop}%
\bibitem [{\citenamefont {{Jeon}}\ \emph {et~al.}(2024)\citenamefont {{Jeon}},
  \citenamefont {{Kim}},\ and\ \citenamefont {{An}}}]{2024arXiv240809486J}%
  \BibitemOpen
  \bibfield  {author} {\bibinfo {author} {\bibfnamefont {M.}~\bibnamefont
  {{Jeon}}}, \bibinfo {author} {\bibfnamefont {J.}~\bibnamefont {{Kim}}},\ and\
  \bibinfo {author} {\bibfnamefont {K.}~\bibnamefont {{An}}},\ }\bibfield
  {title} {\bibinfo {title} {{Numerical analysis of a
  superradiance-sideband-assisted laser with a zero frequency pulling and a
  narrow linewidth}},\ }\href@noop {} {\  (\bibinfo {year} {2024})},\ \Eprint
  {https://arxiv.org/abs/2408.09486} {arXiv:2408.09486 [quant-ph]} \BibitemShut
  {NoStop}%
\bibitem [{\citenamefont {{Nye}}\ and\ \citenamefont
  {{Berry}}(1974)}]{1974RoyalSociety}%
  \BibitemOpen
  \bibfield  {author} {\bibinfo {author} {\bibfnamefont {J.~F.}\ \bibnamefont
  {{Nye}}}\ and\ \bibinfo {author} {\bibfnamefont {M.~V.}\ \bibnamefont
  {{Berry}}},\ }\bibfield  {title} {\bibinfo {title} {Dislocations in wave
  trains},\ }\href {https://doi.org/10.1098/rspa.1974.0012} {\bibfield
  {journal} {\bibinfo  {journal} {Proc. R. Soc. Lond. A}\ }\textbf {\bibinfo
  {volume} {336}},\ \bibinfo {pages} {165} (\bibinfo {year}
  {1974})}\BibitemShut {NoStop}%
\bibitem [{\citenamefont {Volke-Sep\'ulveda}\ \emph {et~al.}(2008)\citenamefont
  {Volke-Sep\'ulveda}, \citenamefont {Santill\'an},\ and\ \citenamefont
  {Boullosa}}]{PhysRevLett.100.024302}%
  \BibitemOpen
  \bibfield  {author} {\bibinfo {author} {\bibfnamefont {K.}~\bibnamefont
  {Volke-Sep\'ulveda}}, \bibinfo {author} {\bibfnamefont {A.~O.}\ \bibnamefont
  {Santill\'an}},\ and\ \bibinfo {author} {\bibfnamefont {R.~R.}\ \bibnamefont
  {Boullosa}},\ }\bibfield  {title} {\bibinfo {title} {Transfer of angular
  momentum to matter from acoustical vortices in free space},\ }\href
  {https://doi.org/10.1103/PhysRevLett.100.024302} {\bibfield  {journal}
  {\bibinfo  {journal} {Phys. Rev. Lett.}\ }\textbf {\bibinfo {volume} {100}},\
  \bibinfo {pages} {024302} (\bibinfo {year} {2008})}\BibitemShut {NoStop}%
\bibitem [{\citenamefont {{Cromb}}\ \emph {et~al.}(2020)\citenamefont
  {{Cromb}}, \citenamefont {{Gibson}}, \citenamefont {{Toninelli}},
  \citenamefont {{Padgett}}, \citenamefont {{Wright}},\ and\ \citenamefont
  {{Faccio}}}]{2020NatPh..16.1069C}%
  \BibitemOpen
  \bibfield  {author} {\bibinfo {author} {\bibfnamefont {M.}~\bibnamefont
  {{Cromb}}}, \bibinfo {author} {\bibfnamefont {G.~M.}\ \bibnamefont
  {{Gibson}}}, \bibinfo {author} {\bibfnamefont {E.}~\bibnamefont
  {{Toninelli}}}, \bibinfo {author} {\bibfnamefont {M.~J.}\ \bibnamefont
  {{Padgett}}}, \bibinfo {author} {\bibfnamefont {E.~M.}\ \bibnamefont
  {{Wright}}},\ and\ \bibinfo {author} {\bibfnamefont {D.}~\bibnamefont
  {{Faccio}}},\ }\bibfield  {title} {\bibinfo {title} {{Amplification of waves
  from a rotating body}},\ }\href {https://doi.org/10.1038/s41567-020-0944-3}
  {\bibfield  {journal} {\bibinfo  {journal} {Nat. Phys.}\ }\textbf {\bibinfo
  {volume} {16}},\ \bibinfo {pages} {1069} (\bibinfo {year}
  {2020})}\BibitemShut {NoStop}%
\bibitem [{\citenamefont {{Gibson}}\ \emph {et~al.}(2018)\citenamefont
  {{Gibson}}, \citenamefont {{Toninelli}}, \citenamefont {{Horsley}},
  \citenamefont {{Spalding}}, \citenamefont {{Hendry}}, \citenamefont
  {{Phillips}},\ and\ \citenamefont {{Padgett}}}]{2018PNAS}%
  \BibitemOpen
  \bibfield  {author} {\bibinfo {author} {\bibfnamefont {G.~M.}\ \bibnamefont
  {{Gibson}}}, \bibinfo {author} {\bibfnamefont {E.}~\bibnamefont
  {{Toninelli}}}, \bibinfo {author} {\bibfnamefont {S.~A.~R.}\ \bibnamefont
  {{Horsley}}}, \bibinfo {author} {\bibfnamefont {G.~C.}\ \bibnamefont
  {{Spalding}}}, \bibinfo {author} {\bibfnamefont {E.}~\bibnamefont
  {{Hendry}}}, \bibinfo {author} {\bibfnamefont {D.~B.}\ \bibnamefont
  {{Phillips}}},\ and\ \bibinfo {author} {\bibfnamefont {M.~J.}\ \bibnamefont
  {{Padgett}}},\ }\bibfield  {title} {\bibinfo {title} {{Reversal of orbital
  angular momentum arising from an extreme Doppler shift}},\ }\href
  {https://doi.org/10.1073/pnas.1720776115} {\bibfield  {journal} {\bibinfo
  {journal} {PNAS}\ }\textbf {\bibinfo {volume} {15}},\ \bibinfo {pages} {3800}
  (\bibinfo {year} {2018})}\BibitemShut {NoStop}%
\bibitem [{\citenamefont {Courtial}\ \emph {et~al.}(1998)\citenamefont
  {Courtial}, \citenamefont {Dholakia}, \citenamefont {Robertson},
  \citenamefont {Allen},\ and\ \citenamefont {Padgett}}]{PhysRevLett.80.3217}%
  \BibitemOpen
  \bibfield  {author} {\bibinfo {author} {\bibfnamefont {J.}~\bibnamefont
  {Courtial}}, \bibinfo {author} {\bibfnamefont {K.}~\bibnamefont {Dholakia}},
  \bibinfo {author} {\bibfnamefont {D.~A.}\ \bibnamefont {Robertson}}, \bibinfo
  {author} {\bibfnamefont {L.}~\bibnamefont {Allen}},\ and\ \bibinfo {author}
  {\bibfnamefont {M.~J.}\ \bibnamefont {Padgett}},\ }\bibfield  {title}
  {\bibinfo {title} {Measurement of the rotational frequency shift imparted to
  a rotating light beam possessing orbital angular momentum},\ }\href
  {https://doi.org/10.1103/PhysRevLett.80.3217} {\bibfield  {journal} {\bibinfo
   {journal} {Phys. Rev. Lett.}\ }\textbf {\bibinfo {volume} {80}},\ \bibinfo
  {pages} {3217} (\bibinfo {year} {1998})}\BibitemShut {NoStop}%
\bibitem [{\citenamefont {{Skeldon}}\ \emph {et~al.}(2008)\citenamefont
  {{Skeldon}}, \citenamefont {{Wilson}}, \citenamefont {{Edgar}},\ and\
  \citenamefont {{Padgett}}}]{Skeldon_2008}%
  \BibitemOpen
  \bibfield  {author} {\bibinfo {author} {\bibfnamefont {K.~D.}\ \bibnamefont
  {{Skeldon}}}, \bibinfo {author} {\bibfnamefont {C.}~\bibnamefont {{Wilson}}},
  \bibinfo {author} {\bibfnamefont {M.}~\bibnamefont {{Edgar}}},\ and\ \bibinfo
  {author} {\bibfnamefont {M.~J.}\ \bibnamefont {{Padgett}}},\ }\bibfield
  {title} {\bibinfo {title} {An acoustic spanner and its associated rotational
  doppler shift},\ }\href {https://doi.org/10.1088/1367-2630/10/1/013018}
  {\bibfield  {journal} {\bibinfo  {journal} {New J. Phys.}\ }\textbf {\bibinfo
  {volume} {10}},\ \bibinfo {pages} {013018} (\bibinfo {year}
  {2008})}\BibitemShut {NoStop}%
\bibitem [{\citenamefont {{Lavery}}\ \emph {et~al.}(2018)\citenamefont
  {{Lavery}}, \citenamefont {{Speirits}}, \citenamefont {{Barnett}},\ and\
  \citenamefont {{Padgett}}}]{2013SCIENCE}%
  \BibitemOpen
  \bibfield  {author} {\bibinfo {author} {\bibfnamefont {M.~P.~J.}\
  \bibnamefont {{Lavery}}}, \bibinfo {author} {\bibfnamefont {F.~C.}\
  \bibnamefont {{Speirits}}}, \bibinfo {author} {\bibfnamefont {S.~M.}\
  \bibnamefont {{Barnett}}},\ and\ \bibinfo {author} {\bibfnamefont {M.~J.}\
  \bibnamefont {{Padgett}}},\ }\bibfield  {title} {\bibinfo {title} {{Detection
  of a spinning object using light's orbital angular momentum}},\ }\href
  {https://doi.org/10.1126/science.1239936} {\bibfield  {journal} {\bibinfo
  {journal} {Science}\ }\textbf {\bibinfo {volume} {341}},\ \bibinfo {pages}
  {537} (\bibinfo {year} {2018})}\BibitemShut {NoStop}%
\bibitem [{\citenamefont {{Rosales-Guzmán}}\ \emph {et~al.}(2013)\citenamefont
  {{Rosales-Guzmán}}, \citenamefont {{Hermosa}}, \citenamefont {{Belmonte}},\
  and\ \citenamefont {{Torres}}}]{20131002ScientificReports}%
  \BibitemOpen
  \bibfield  {author} {\bibinfo {author} {\bibfnamefont {C.}~\bibnamefont
  {{Rosales-Guzmán}}}, \bibinfo {author} {\bibfnamefont {N.}~\bibnamefont
  {{Hermosa}}}, \bibinfo {author} {\bibfnamefont {A.}~\bibnamefont
  {{Belmonte}}},\ and\ \bibinfo {author} {\bibfnamefont {J.~P.}\ \bibnamefont
  {{Torres}}},\ }\bibfield  {title} {\bibinfo {title} {{Dislocations in wave
  trains}},\ }\href {https://doi.org/10.1038/srep02815} {\bibfield  {journal}
  {\bibinfo  {journal} {Sci. Rep.}\ }\textbf {\bibinfo {volume} {3}},\ \bibinfo
  {pages} {2815} (\bibinfo {year} {2013})}\BibitemShut {NoStop}%
\bibitem [{\citenamefont {Phillips}\ \emph {et~al.}(2014)\citenamefont
  {Phillips}, \citenamefont {Lee}, \citenamefont {Speirits}, \citenamefont
  {Barnett}, \citenamefont {Simpson}, \citenamefont {Lavery}, \citenamefont
  {Padgett},\ and\ \citenamefont {Gibson}}]{PhysRevA.90.011801}%
  \BibitemOpen
  \bibfield  {author} {\bibinfo {author} {\bibfnamefont {D.~B.}\ \bibnamefont
  {Phillips}}, \bibinfo {author} {\bibfnamefont {M.~P.}\ \bibnamefont {Lee}},
  \bibinfo {author} {\bibfnamefont {F.~C.}\ \bibnamefont {Speirits}}, \bibinfo
  {author} {\bibfnamefont {S.~M.}\ \bibnamefont {Barnett}}, \bibinfo {author}
  {\bibfnamefont {S.~H.}\ \bibnamefont {Simpson}}, \bibinfo {author}
  {\bibfnamefont {M.~P.~J.}\ \bibnamefont {Lavery}}, \bibinfo {author}
  {\bibfnamefont {M.~J.}\ \bibnamefont {Padgett}},\ and\ \bibinfo {author}
  {\bibfnamefont {G.~M.}\ \bibnamefont {Gibson}},\ }\bibfield  {title}
  {\bibinfo {title} {Rotational doppler velocimetry to probe the angular
  velocity of spinning microparticles},\ }\href
  {https://doi.org/10.1103/PhysRevA.90.011801} {\bibfield  {journal} {\bibinfo
  {journal} {Phys. Rev. A}\ }\textbf {\bibinfo {volume} {90}},\ \bibinfo
  {pages} {011801} (\bibinfo {year} {2014})}\BibitemShut {NoStop}%
\bibitem [{\citenamefont {{Teukolsky}}\ and\ \citenamefont
  {{Press}}(1974)}]{1974ApJ...193..443T}%
  \BibitemOpen
  \bibfield  {author} {\bibinfo {author} {\bibfnamefont {S.~A.}\ \bibnamefont
  {{Teukolsky}}}\ and\ \bibinfo {author} {\bibfnamefont {W.~H.}\ \bibnamefont
  {{Press}}},\ }\bibfield  {title} {\bibinfo {title} {{Perturbations of a
  rotating black hole. III. Interaction of the hole with gravitational and
  electromagnetic radiation.}},\ }\href {https://doi.org/10.1086/153180}
  {\bibfield  {journal} {\bibinfo  {journal} {\apj}\ }\textbf {\bibinfo
  {volume} {193}},\ \bibinfo {pages} {443} (\bibinfo {year}
  {1974})}\BibitemShut {NoStop}%
\bibitem [{\citenamefont {{Press}}\ and\ \citenamefont
  {{Teukolsky}}(1972)}]{1972PRE}%
  \BibitemOpen
  \bibfield  {author} {\bibinfo {author} {\bibfnamefont {W.}~\bibnamefont
  {{Press}}}\ and\ \bibinfo {author} {\bibfnamefont {S.}~\bibnamefont
  {{Teukolsky}}},\ }\bibfield  {title} {\bibinfo {title} {{Floating Orbits,
  Superradiant Scattering and the Black-hole Bomb.}},\ }\href
  {https://doi.org/10.1038/238211a0} {\bibfield  {journal} {\bibinfo  {journal}
  {Nature}\ }\textbf {\bibinfo {volume} {238}},\ \bibinfo {pages} {211}
  (\bibinfo {year} {1972})}\BibitemShut {NoStop}%
\bibitem [{\citenamefont {Cardoso}\ \emph
  {et~al.}(2004{\natexlab{a}})\citenamefont {Cardoso}, \citenamefont {Dias},
  \citenamefont {Lemos},\ and\ \citenamefont {Yoshida}}]{Cardoso:2004nk}%
  \BibitemOpen
  \bibfield  {author} {\bibinfo {author} {\bibfnamefont {V.}~\bibnamefont
  {Cardoso}}, \bibinfo {author} {\bibfnamefont {O.~J.~C.}\ \bibnamefont
  {Dias}}, \bibinfo {author} {\bibfnamefont {J.~P.~S.}\ \bibnamefont {Lemos}},\
  and\ \bibinfo {author} {\bibfnamefont {S.}~\bibnamefont {Yoshida}},\
  }\bibfield  {title} {\bibinfo {title} {{The Black hole bomb and superradiant
  instabilities}},\ }\href {https://doi.org/10.1103/PhysRevD.70.049903}
  {\bibfield  {journal} {\bibinfo  {journal} {Phys. Rev. D}\ }\textbf {\bibinfo
  {volume} {70}},\ \bibinfo {pages} {044039} (\bibinfo {year}
  {2004}{\natexlab{a}})},\ \Eprint {https://arxiv.org/abs/hep-th/0404096}
  {arXiv:hep-th/0404096} \BibitemShut {NoStop}%
\bibitem [{\citenamefont {{Pani}}\ \emph {et~al.}(2012)\citenamefont {{Pani}},
  \citenamefont {{Cardoso}}, \citenamefont {{Gualtieri}}, \citenamefont
  {{Berti}},\ and\ \citenamefont {{Ishibashi}}}]{2012PhRvL.109m1102P}%
  \BibitemOpen
  \bibfield  {author} {\bibinfo {author} {\bibfnamefont {P.}~\bibnamefont
  {{Pani}}}, \bibinfo {author} {\bibfnamefont {V.}~\bibnamefont {{Cardoso}}},
  \bibinfo {author} {\bibfnamefont {L.}~\bibnamefont {{Gualtieri}}}, \bibinfo
  {author} {\bibfnamefont {E.}~\bibnamefont {{Berti}}},\ and\ \bibinfo {author}
  {\bibfnamefont {A.}~\bibnamefont {{Ishibashi}}},\ }\bibfield  {title}
  {\bibinfo {title} {{Black-Hole Bombs and Photon-Mass Bounds}},\ }\href
  {https://doi.org/10.1103/PhysRevLett.109.131102} {\bibfield  {journal}
  {\bibinfo  {journal} {\prl}\ }\textbf {\bibinfo {volume} {109}},\ \bibinfo
  {eid} {131102} (\bibinfo {year} {2012})}\BibitemShut {NoStop}%
\bibitem [{\citenamefont {{Degollado}}\ and\ \citenamefont
  {{Herdeiro}}(2014)}]{2014PhRvD..89f3005D}%
  \BibitemOpen
  \bibfield  {author} {\bibinfo {author} {\bibfnamefont {J.~C.}\ \bibnamefont
  {{Degollado}}}\ and\ \bibinfo {author} {\bibfnamefont {C.~A.~R.}\
  \bibnamefont {{Herdeiro}}},\ }\bibfield  {title} {\bibinfo {title} {{Time
  evolution of superradiant instabilities for charged black holes in a
  cavity}},\ }\href {https://doi.org/10.1103/PhysRevD.89.063005} {\bibfield
  {journal} {\bibinfo  {journal} {\prd}\ }\textbf {\bibinfo {volume} {89}},\
  \bibinfo {eid} {063005} (\bibinfo {year} {2014})}\BibitemShut {NoStop}%
\bibitem [{\citenamefont {{Dias}}\ and\ \citenamefont
  {{Masachs}}(2018)}]{2018CQGra..35r4001D}%
  \BibitemOpen
  \bibfield  {author} {\bibinfo {author} {\bibfnamefont {O.~J.~C.}\
  \bibnamefont {{Dias}}}\ and\ \bibinfo {author} {\bibfnamefont
  {R.}~\bibnamefont {{Masachs}}},\ }\bibfield  {title} {\bibinfo {title}
  {{Charged black hole bombs in a Minkowski cavity}},\ }\href
  {https://doi.org/10.1088/1361-6382/aad70b} {\bibfield  {journal} {\bibinfo
  {journal} {Class. Quant. Grav.}\ }\textbf {\bibinfo {volume} {35}},\ \bibinfo
  {eid} {184001} (\bibinfo {year} {2018})}\BibitemShut {NoStop}%
\bibitem [{\citenamefont {{Hod}}(2013)}]{2013PhRvD..88f4055H}%
  \BibitemOpen
  \bibfield  {author} {\bibinfo {author} {\bibfnamefont {S.}~\bibnamefont
  {{Hod}}},\ }\bibfield  {title} {\bibinfo {title} {{Analytic treatment of the
  charged black-hole-mirror bomb in the highly explosive regime}},\ }\href
  {https://doi.org/10.1103/PhysRevD.88.064055} {\bibfield  {journal} {\bibinfo
  {journal} {\prd}\ }\textbf {\bibinfo {volume} {88}},\ \bibinfo {eid} {064055}
  (\bibinfo {year} {2013})}\BibitemShut {NoStop}%
\bibitem [{\citenamefont {{Li}}\ \emph {et~al.}(2015)\citenamefont {{Li}},
  \citenamefont {{Zhao}},\ and\ \citenamefont {{Zhang}}}]{2015CoTPh..63..569L}%
  \BibitemOpen
  \bibfield  {author} {\bibinfo {author} {\bibfnamefont {R.}~\bibnamefont
  {{Li}}}, \bibinfo {author} {\bibfnamefont {J.~K.}\ \bibnamefont {{Zhao}}},\
  and\ \bibinfo {author} {\bibfnamefont {Y.~M.}\ \bibnamefont {{Zhang}}},\
  }\bibfield  {title} {\bibinfo {title} {{Superradiant Instability of
  D-Dimensional Reissner{\textemdash}Nordstr{\"o}m Black Hole Mirror System}},\
  }\href {https://doi.org/10.1088/0253-6102/63/5/569} {\bibfield  {journal}
  {\bibinfo  {journal} {Commun. Theor. Phys.}\ }\textbf {\bibinfo {volume}
  {63}},\ \bibinfo {eid} {569} (\bibinfo {year} {2015})}\BibitemShut {NoStop}%
\bibitem [{\citenamefont {{Visser}}(1998)}]{MattVisser1998}%
  \BibitemOpen
  \bibfield  {author} {\bibinfo {author} {\bibfnamefont {M.}~\bibnamefont
  {{Visser}}},\ }\bibfield  {title} {\bibinfo {title} {Acoustic black holes:
  horizons, ergospheres and hawking radiation},\ }\href
  {https://doi.org/10.1088/0264-9381/15/6/024} {\bibfield  {journal} {\bibinfo
  {journal} {Class. Quant. Grav.}\ }\textbf {\bibinfo {volume} {15}},\ \bibinfo
  {pages} {1767} (\bibinfo {year} {1998})}\BibitemShut {NoStop}%
\bibitem [{\citenamefont {Berti}\ \emph {et~al.}(2004)\citenamefont {Berti},
  \citenamefont {Cardoso},\ and\ \citenamefont {Lemos}}]{Berti:2004ju}%
  \BibitemOpen
  \bibfield  {author} {\bibinfo {author} {\bibfnamefont {E.}~\bibnamefont
  {Berti}}, \bibinfo {author} {\bibfnamefont {V.}~\bibnamefont {Cardoso}},\
  and\ \bibinfo {author} {\bibfnamefont {J.~P.~S.}\ \bibnamefont {Lemos}},\
  }\bibfield  {title} {\bibinfo {title} {{Quasinormal modes and classical wave
  propagation in analogue black holes}},\ }\href
  {https://doi.org/10.1103/PhysRevD.70.124006} {\bibfield  {journal} {\bibinfo
  {journal} {Phys. Rev. D}\ }\textbf {\bibinfo {volume} {70}},\ \bibinfo
  {pages} {124006} (\bibinfo {year} {2004})},\ \Eprint
  {https://arxiv.org/abs/gr-qc/0408099} {arXiv:gr-qc/0408099} \BibitemShut
  {NoStop}%
\bibitem [{\citenamefont {Cardoso}\ \emph
  {et~al.}(2004{\natexlab{b}})\citenamefont {Cardoso}, \citenamefont {Lemos},\
  and\ \citenamefont {Yoshida}}]{Cardoso:2004fi}%
  \BibitemOpen
  \bibfield  {author} {\bibinfo {author} {\bibfnamefont {V.}~\bibnamefont
  {Cardoso}}, \bibinfo {author} {\bibfnamefont {J.~P.~S.}\ \bibnamefont
  {Lemos}},\ and\ \bibinfo {author} {\bibfnamefont {S.}~\bibnamefont
  {Yoshida}},\ }\bibfield  {title} {\bibinfo {title} {{Quasinormal modes and
  stability of the rotating acoustic black hole: Numerical analysis}},\ }\href
  {https://doi.org/10.1103/PhysRevD.70.124032} {\bibfield  {journal} {\bibinfo
  {journal} {Phys. Rev. D}\ }\textbf {\bibinfo {volume} {70}},\ \bibinfo
  {pages} {124032} (\bibinfo {year} {2004}{\natexlab{b}})},\ \Eprint
  {https://arxiv.org/abs/gr-qc/0410107} {arXiv:gr-qc/0410107} \BibitemShut
  {NoStop}%
\bibitem [{\citenamefont {Lepe}\ and\ \citenamefont
  {Saavedra}(2005)}]{Lepe:2004kv}%
  \BibitemOpen
  \bibfield  {author} {\bibinfo {author} {\bibfnamefont {S.}~\bibnamefont
  {Lepe}}\ and\ \bibinfo {author} {\bibfnamefont {J.}~\bibnamefont
  {Saavedra}},\ }\bibfield  {title} {\bibinfo {title} {{Quasinormal modes,
  superradiance and area spectrum for 2+1 acoustic black holes}},\ }\href
  {https://doi.org/10.1016/j.physletb.2005.05.021} {\bibfield  {journal}
  {\bibinfo  {journal} {Phys. Lett. B}\ }\textbf {\bibinfo {volume} {617}},\
  \bibinfo {pages} {174} (\bibinfo {year} {2005})},\ \Eprint
  {https://arxiv.org/abs/gr-qc/0410074} {arXiv:gr-qc/0410074} \BibitemShut
  {NoStop}%
\bibitem [{\citenamefont {Mironov}(1988)}]{1988MIR}%
  \BibitemOpen
  \bibfield  {author} {\bibinfo {author} {\bibfnamefont {M.}~\bibnamefont
  {Mironov}},\ }\bibfield  {title} {\bibinfo {title} {Propagation of a flexural
  wave in a plate whose thickness decreases smoothly to zero in a finite
  interval},\ }\href@noop {} {\bibfield  {journal} {\bibinfo  {journal} {Sov.
  Phys. Acoust.}\ }\textbf {\bibinfo {volume} {34}},\ \bibinfo {pages} {318}
  (\bibinfo {year} {1988})}\BibitemShut {NoStop}%
\bibitem [{\citenamefont {{Hawking}}(1974)}]{1974HAW}%
  \BibitemOpen
  \bibfield  {author} {\bibinfo {author} {\bibfnamefont {S.~W.}\ \bibnamefont
  {{Hawking}}},\ }\bibfield  {title} {\bibinfo {title} {{Black hole
  explosions?}},\ }\href {https://doi.org/10.1038/248030a0} {\bibfield
  {journal} {\bibinfo  {journal} {Nature}\ }\textbf {\bibinfo {volume} {248}},\
  \bibinfo {pages} {30} (\bibinfo {year} {1974})}\BibitemShut {NoStop}%
\bibitem [{\citenamefont {{Hawking}}(1976)}]{1976HAWW}%
  \BibitemOpen
  \bibfield  {author} {\bibinfo {author} {\bibfnamefont {S.~W.}\ \bibnamefont
  {{Hawking}}},\ }\bibfield  {title} {\bibinfo {title} {Particle creation by
  black holes},\ }\href {https://doi.org/10.1007/BF01608497} {\bibfield
  {journal} {\bibinfo  {journal} {Commun. math Phys.}\ }\textbf {\bibinfo
  {volume} {46}},\ \bibinfo {pages} {206} (\bibinfo {year} {1976})}\BibitemShut
  {NoStop}%
\bibitem [{\citenamefont {Hawking}(1976)}]{PhysRevD.14.2460}%
  \BibitemOpen
  \bibfield  {author} {\bibinfo {author} {\bibfnamefont {S.~W.}\ \bibnamefont
  {Hawking}},\ }\bibfield  {title} {\bibinfo {title} {Breakdown of
  predictability in gravitational collapse},\ }\href
  {https://doi.org/10.1103/PhysRevD.14.2460} {\bibfield  {journal} {\bibinfo
  {journal} {Phys. Rev. D}\ }\textbf {\bibinfo {volume} {14}},\ \bibinfo
  {pages} {2460} (\bibinfo {year} {1976})}\BibitemShut {NoStop}%
\bibitem [{\citenamefont {{Parikh}}\ and\ \citenamefont
  {{Wilczek}}(2000)}]{2000PhRvL..85.5042P}%
  \BibitemOpen
  \bibfield  {author} {\bibinfo {author} {\bibfnamefont {M.~K.}\ \bibnamefont
  {{Parikh}}}\ and\ \bibinfo {author} {\bibfnamefont {F.}~\bibnamefont
  {{Wilczek}}},\ }\bibfield  {title} {\bibinfo {title} {{Hawking Radiation As
  Tunneling}},\ }\href {https://doi.org/10.1103/PhysRevLett.85.5042} {\bibfield
   {journal} {\bibinfo  {journal} {\prl}\ }\textbf {\bibinfo {volume} {85}},\
  \bibinfo {pages} {5042} (\bibinfo {year} {2000})}\BibitemShut {NoStop}%
\bibitem [{\citenamefont {{Hemming}}\ and\ \citenamefont
  {{Keski-Vakkuri}}(2001)}]{2001PhRvD..64d4006H}%
  \BibitemOpen
  \bibfield  {author} {\bibinfo {author} {\bibfnamefont {S.}~\bibnamefont
  {{Hemming}}}\ and\ \bibinfo {author} {\bibfnamefont {E.}~\bibnamefont
  {{Keski-Vakkuri}}},\ }\bibfield  {title} {\bibinfo {title} {{Hawking
  radiation from AdS black holes}},\ }\href
  {https://doi.org/10.1103/PhysRevD.64.044006} {\bibfield  {journal} {\bibinfo
  {journal} {\prd}\ }\textbf {\bibinfo {volume} {64}},\ \bibinfo {eid} {044006}
  (\bibinfo {year} {2001})}\BibitemShut {NoStop}%
\bibitem [{\citenamefont {{Krylov}}(1989)}]{1989KVV}%
  \BibitemOpen
  \bibfield  {author} {\bibinfo {author} {\bibfnamefont {V.~V.}\ \bibnamefont
  {{Krylov}}},\ }\bibfield  {title} {\bibinfo {title} {{Surface effects and
  surface acoustic waves.}},\ }\href@noop {} {\bibfield  {journal} {\bibinfo
  {journal} {Prog. Surf. Sci.}\ }\textbf {\bibinfo {volume} {32}},\ \bibinfo
  {pages} {39} (\bibinfo {year} {1989})}\BibitemShut {NoStop}%
\bibitem [{\citenamefont {{Krylov}}\ and\ \citenamefont
  {{Tilman}}(2004)}]{KRYLOV2004605}%
  \BibitemOpen
  \bibfield  {author} {\bibinfo {author} {\bibfnamefont {V.~V.}\ \bibnamefont
  {{Krylov}}}\ and\ \bibinfo {author} {\bibfnamefont {F.~J. B.~S.}\
  \bibnamefont {{Tilman}}},\ }\bibfield  {title} {\bibinfo {title} {Acoustic
  ‘black holes’ for flexural waves as effective vibration dampers},\ }\href
  {https://doi.org/https://doi.org/10.1016/j.jsv.2003.05.010} {\bibfield
  {journal} {\bibinfo  {journal} {J. Sound Vib.}\ }\textbf {\bibinfo {volume}
  {274}},\ \bibinfo {pages} {605} (\bibinfo {year} {2004})}\BibitemShut
  {NoStop}%
\bibitem [{\citenamefont {Denis}\ \emph {et~al.}(2014)\citenamefont {Denis},
  \citenamefont {Pelat}, \citenamefont {Gautier},\ and\ \citenamefont
  {Elie}}]{DENIS20142475}%
  \BibitemOpen
  \bibfield  {author} {\bibinfo {author} {\bibfnamefont {V.}~\bibnamefont
  {Denis}}, \bibinfo {author} {\bibfnamefont {A.}~\bibnamefont {Pelat}},
  \bibinfo {author} {\bibfnamefont {F.}~\bibnamefont {Gautier}},\ and\ \bibinfo
  {author} {\bibfnamefont {B.}~\bibnamefont {Elie}},\ }\bibfield  {title}
  {\bibinfo {title} {Modal overlap factor of a beam with an acoustic black hole
  termination},\ }\href
  {https://doi.org/https://doi.org/10.1016/j.jsv.2014.02.005} {\bibfield
  {journal} {\bibinfo  {journal} {J. Sound Vib.}\ }\textbf {\bibinfo {volume}
  {333}},\ \bibinfo {pages} {2475} (\bibinfo {year} {2014})}\BibitemShut
  {NoStop}%
\bibitem [{\citenamefont {Tang}\ and\ \citenamefont
  {Cheng}(2017)}]{TANG2017116}%
  \BibitemOpen
  \bibfield  {author} {\bibinfo {author} {\bibfnamefont {L.}~\bibnamefont
  {Tang}}\ and\ \bibinfo {author} {\bibfnamefont {L.}~\bibnamefont {Cheng}},\
  }\bibfield  {title} {\bibinfo {title} {Enhanced acoustic black hole effect in
  beams with a modified thickness profile and extended platform},\ }\href
  {https://doi.org/https://doi.org/10.1016/j.jsv.2016.11.010} {\bibfield
  {journal} {\bibinfo  {journal} {J. Sound Vib.}\ }\textbf {\bibinfo {volume}
  {391}},\ \bibinfo {pages} {116} (\bibinfo {year} {2017})}\BibitemShut
  {NoStop}%
\bibitem [{\citenamefont {Pelat}\ \emph {et~al.}(2020)\citenamefont {Pelat},
  \citenamefont {Gautier}, \citenamefont {Conlon},\ and\ \citenamefont
  {Semperlotti}}]{PELAT2020115316}%
  \BibitemOpen
  \bibfield  {author} {\bibinfo {author} {\bibfnamefont {A.}~\bibnamefont
  {Pelat}}, \bibinfo {author} {\bibfnamefont {F.}~\bibnamefont {Gautier}},
  \bibinfo {author} {\bibfnamefont {S.~C.}\ \bibnamefont {Conlon}},\ and\
  \bibinfo {author} {\bibfnamefont {F.}~\bibnamefont {Semperlotti}},\
  }\bibfield  {title} {\bibinfo {title} {The acoustic black hole: A review of
  theory and applications},\ }\href
  {https://doi.org/https://doi.org/10.1016/j.jsv.2020.115316} {\bibfield
  {journal} {\bibinfo  {journal} {J. Sound Vib.}\ }\textbf {\bibinfo {volume}
  {476}},\ \bibinfo {pages} {115316} (\bibinfo {year} {2020})}\BibitemShut
  {NoStop}%
\bibitem [{\citenamefont {Cheli}\ and\ \citenamefont
  {Diana}(2015)}]{Cheli2015}%
  \BibitemOpen
  \bibfield  {author} {\bibinfo {author} {\bibfnamefont {F.}~\bibnamefont
  {Cheli}}\ and\ \bibinfo {author} {\bibfnamefont {G.}~\bibnamefont {Diana}},\
  }\bibinfo {title} {Introduction to the finite element method},\ in\ \href
  {https://doi.org/10.1007/978-3-319-18200-1_4} {\emph {\bibinfo {booktitle}
  {Advanced Dynamics of Mechanical Systems}}}\ (\bibinfo  {publisher} {Springer
  International Publishing},\ \bibinfo {address} {Cham},\ \bibinfo {year}
  {2015})\ pp.\ \bibinfo {pages} {311--412}\BibitemShut {NoStop}%
\bibitem [{\citenamefont {Zienkiewicz}\ \emph {et~al.}(2005)\citenamefont
  {Zienkiewicz}, \citenamefont {Taylor},\ and\ \citenamefont
  {Zhu}}]{Zienkiewicz2005TheFE}%
  \BibitemOpen
  \bibfield  {author} {\bibinfo {author} {\bibfnamefont {O.~C.}\ \bibnamefont
  {Zienkiewicz}}, \bibinfo {author} {\bibfnamefont {R.~L.}\ \bibnamefont
  {Taylor}},\ and\ \bibinfo {author} {\bibfnamefont {J.~Z.}\ \bibnamefont
  {Zhu}},\ }\bibfield  {title} {\bibinfo {title} {The finite element method:
  Its basis and fundamentals}\ }(\bibinfo {year} {2005})\BibitemShut {NoStop}%
\bibitem [{\citenamefont {COMSOL}()}]{2024comsol}%
  \BibitemOpen
  \bibfield  {author} {\bibinfo {author} {\bibnamefont {COMSOL}},\ }\href
  {https://www.comsol.com} {\bibinfo  {journal} {COMSOL Multiphysics Modeling
  Software}\ }\BibitemShut {NoStop}%
\bibitem [{\citenamefont {Pierce}\ and\ \citenamefont
  {Beyer}(1990)}]{10.1121/1.399390}%
  \BibitemOpen
\bibfield  {journal} {  }\bibfield  {author} {\bibinfo {author} {\bibfnamefont
  {A.~D.}\ \bibnamefont {Pierce}}\ and\ \bibinfo {author} {\bibfnamefont
  {R.~T.}\ \bibnamefont {Beyer}},\ }\bibfield  {title} {\bibinfo {title}
  {{Acoustics: An Introduction to Its Physical Principles
  and Applications}},\ }\href {https://doi.org/10.1121/1.399390} {\bibfield
  {journal} {\bibinfo  {journal} {J. Acoust. Soc. Am.}\ }\textbf {\bibinfo
  {volume} {87}},\ \bibinfo {pages} {1826} (\bibinfo {year}
  {1990})}\BibitemShut {NoStop}%
\bibitem [{\citenamefont {{Rienstra}}\ and\ \citenamefont
  {{Hirschberg}}()}]{2021SWRA}%
  \BibitemOpen
  \bibfield  {author} {\bibinfo {author} {\bibfnamefont {S.~W.}\ \bibnamefont
  {{Rienstra}}}\ and\ \bibinfo {author} {\bibfnamefont {A.}~\bibnamefont
  {{Hirschberg}}},\ }\href {https://sjoerdr.win.tue.nl/papers/boek.pdf}
  {\bibinfo  {journal} {An Introduction to Acoustics}\ }\BibitemShut {NoStop}%
\bibitem [{\citenamefont {Delany}\ and\ \citenamefont
  {Bazley}(1970)}]{1970DEM}%
  \BibitemOpen
\bibfield  {journal} {  }\bibfield  {author} {\bibinfo {author} {\bibfnamefont
  {M.~E.}\ \bibnamefont {Delany}}\ and\ \bibinfo {author} {\bibfnamefont
  {E.~N.}\ \bibnamefont {Bazley}},\ }\bibfield  {title} {\bibinfo {title}
  {{Acoustical properties of fibrous absorbent materials}},\ }\href
  {https://doi.org/10.1016/0003-682X(70)90031-9} {\bibfield  {journal}
  {\bibinfo  {journal} {Appl. Acoust.}\ }\textbf {\bibinfo {volume} {3}},\
  \bibinfo {pages} {105} (\bibinfo {year} {1970})}\BibitemShut {NoStop}%
\bibitem [{\citenamefont {Yasushi}(1990)}]{199019}%
  \BibitemOpen
  \bibfield  {author} {\bibinfo {author} {\bibfnamefont {M.}~\bibnamefont
  {Yasushi}},\ }\bibfield  {title} {\bibinfo {title} {Acoustical properties of
  porous materials-modifications of delany-bazley models-},\ }\href
  {https://doi.org/10.1250/ast.11.19} {\bibfield  {journal} {\bibinfo
  {journal} {J. Acoust. Soc. Jpn. (E)}\ }\textbf {\bibinfo {volume} {11}},\
  \bibinfo {pages} {19} (\bibinfo {year} {1990})}\BibitemShut {NoStop}%
\bibitem [{\citenamefont {Fridolin}(2004)}]{Mechel2004FormulasOA}%
  \BibitemOpen
  \bibfield  {author} {\bibinfo {author} {\bibfnamefont {P.~M.}\ \bibnamefont
  {Fridolin}},\ }\href {https://api.semanticscholar.org/CorpusID:109080228}
  {\bibfield  {journal} {\bibinfo  {journal} {Formulas of Acoustics.}\ }
  (\bibinfo {year} {2004})}\BibitemShut {NoStop}%
\end{thebibliography}
\end{document}